\documentclass{aastex631}

\usepackage{multirow}
\usepackage{amsmath}
\usepackage{hhline}
\usepackage{array, makecell}
\usepackage{booktabs, tabularx}
\usepackage{comment}
\usepackage{rotating}
\usepackage{enumitem}
\usepackage{mathrsfs}

\usepackage{float}
\usepackage{xcolor}
\newcounter{tr}
\ifnum \value{tr}>5
\newcommand{\deletedD}[1]{{\color{red} Damien - Deleted: } \sout{#1}}

\newcommand{\authorcommentD}[1]{{\color{blue} Damien - Comment :} {\color{blue} #1}}

\else
\newcommand{\deletedD}[1]{}

\newcommand{\authorcommentD}[1]{}

\fi
\begin{document}

%%%%%%%%%%%%%%%%%%%%%%%%%%%%%%%%%%%%%%%%%%%%%%%%%%%%%%%%%%
%%%%%%%% TITLE
%%%%%%%%%%%%%%%%%%%%%%%%%%%%%%%%%%%%%%%%%%%%%%%%%%%%%%%%%%
\title{Early--Narrow and Late--Wide Populations of GRB X-ray Flares}

\shorttitle{Two Populations of X-ray Flares}

%%%%%%%%%%%%%%%%%%%%%%%%%%%%%%%%%%%%%%%%%%%%%%%%%%%%%%%%%%
%%%%%%%% AUTHORS
%%%%%%%%%%%%%%%%%%%%%%%%%%%%%%%%%%%%%%%%%%%%%%%%%%%%%%%%%%

\shortauthors{Dereli-B\'egu\'e et al.}

\author[0000-0002-8852-7530]{H. Dereli-B\'egu\'e}
\affiliation{Department of Physics, Bar-Ilan University, Ramat-Gan 52900, Israel}

\author[0000-0001-8667-0889]{A. Pe{'}er}
\affiliation{Department of Physics, Bar-Ilan University, Ramat-Gan 52900, Israel}

%%%%%%%%%%%%%%%%%%%%%%%%%%%%%%%%%%%%%%%%%%%%%%%%%%%%%%%%%%
%%%%%%%% ABSTRACT
%%%%%%%%%%%%%%%%%%%%%%%%%%%%%%%%%%%%%%%%%%%%%%%%%%%%%%%%%%

\begin{abstract}

The physical origin of gamma-ray burst (GRB) X-ray flares remains debated, particularly whether they arise from prolonged central-engine activity or from variability in the external shock. Building on our previous analysis of 89 Swift GRBs with measured redshifts, we investigate the rest-frame temporal properties of X-ray flares detected in 61 bursts. For that, we apply a multivariate Gaussian Mixture Model to the two-dimensional distribution of the rest-frame flare peak time and width.
We identify two statistically distinct populations: early--narrow flares (80\%), with characteristic peak times of approximately $\langle t_{\rm pk} \rangle  \simeq  10^2$~s and $w/t_{\rm pk}\lesssim 1$, and late--wide flares (20\%), with $\langle t_{\rm pk} \rangle \simeq 10^{3.5}$~s and $w/t_{\rm pk}\gtrsim2$.
Of the 61 flares identified as single or first flares in their light curves, 55 belong to the early--narrow population and only 6 to the late--wide population. Of the 36 second flares, 23 (64\%) belong to the early--narrow population, with shorter waiting times, whereas 13 (36\%) belong to the late--wide population, with substantially longer waiting times.
The width--peak-time correlation remains significant for early–narrow flares but not for late--wide flares.
Furthermore, we find no correlation between flare width and energy. We find a decline of flare luminosity with peak time, driven primarily by the early--narrow population, but not statistically significant for late--wide flares. These results imply that a simple shift in time, potentially associated with later episodes of central-engine activity, cannot explain the distinction between the two flare populations. On the other hand, the peak times are consistent with expectations from an accretion disk formed following the  collapse of a stellar-size massive star. 
Combined with our previous finding that flare properties are independent of the presence of an X-ray plateau, these results favor prolonged accretion-driven central-engine activity over external-shock variability as the primary origin of X-ray flares.
\end{abstract}


%%%%%%%%%%%%%%%%%%%%%%%%%%%%%%%%%%%%%%%%%%%%%%%%%%%%%%%%%%
%%%%%%%% INTRODUCTION
%%%%%%%%%%%%%%%%%%%%%%%%%%%%%%%%%%%%%%%%%%%%%%%%%%%%%%%%%%

\section{Introduction} 
\label{sec:intro}

Gamma-ray bursts (GRBs) typically consist of two emission phases: the prompt burst of $\gamma$-rays and a longer-lasting afterglow. The prompt phase shows strong variability, whereas the afterglow is generally smoother and less variable, occasionally  interrupted by X-ray flares whose origin and connection to the prompt emission remain a subject of active investigation. Following the launch of the Neil Gehrels \textit{Swift} Observatory (hereafter \textit{Swift}; \citealt{Gehrels+04}), systematic X-ray afterglow observations revealed that these flares appear as sudden, high-amplitude flux variations during the afterglow. 

X-ray flares are typically observed between $10^2$ and $10^5$ seconds after the prompt emission. They appear in roughly half of all GRBs, predominantly in long bursts, and only rarely in short ones \citep{Burrows+05, Falcone+06, Falcone+07, Chincarini+07, Curran+08, Chincarini+10, Margutti+2011, Bernardini+11}. Most GRBs exhibit one or two flares, with multiple flares being comparatively rare \citep[e.g.,][]{Perri+07, Abdo+11}. Observationally, flares display a broad diversity in their temporal properties, such as peak time $t_{\rm pk}$ ranging from early to late flares or flare width, $w$ being narrow and wide \citep{Chincarini+10, Margutti+2011}.

The origin of X-ray flares and their connection to the prompt phase remain open questions. Observationally, flares share several properties with prompt pulses, such as: (i) hard spectra \citep{Burrows+05, Romano+06, Falcone+06}; (ii) hard-to-soft spectral evolution \citep{Falcone+07}; (iii) similar intensity ratios between successive emission episodes \citep{Chincarini+07}; (iv) a positive correlation between spectral lag and luminosity \citep{Margutti+10}; (v) a correlation between pulse rise times and the minimum variability timescale \citep{Sonbas+13}; and (vi) a power-law-tailed waiting-time distribution extending from prompt $\gamma$-ray pulses to late X-ray flares, up to $\sim 10^3$~s consistent with a common time-dependent Poisson process \citep{Guidorzi+15} \footnote{Note that in that sample, a cutoff of $\sim 10^3$~s for GRBs with known redshift was taken, due to lack of sufficient data at later times.}. The latter suggests that prompt pulses and X-ray flares may represent different manifestations of the same underlying stochastic central-engine activity.

Despite these similarities, there are notable differences between prompt pulses and flares. The number of flares rarely matches the number of prompt pulses, and flare widths increase approximately linearly with time, whereas prompt pulses are narrower and do not exhibit this temporal broadening \citep{Chincarini+07, Chincarini+10, Margutti+10}. Additionally, \citet{SIM23} reported that flare durations correlate positively with their minimum variability timescale and negatively with peak flux, a trend absent in the prompt emission. Moreover, the late prompt pulses appear longer, more asymmetric, and more widely separated than early ones \citep[e.g.,][]{Hakkila+24, Gowri+25}, further highlighting both connections and distinctions between the two emission phases. These differences indicate that flares are not simply a late extension of the prompt emission and motivate detailed comparisons to constrain their physical origin.

Theoretically, a widely accepted interpretation is that X-ray flares are powered by late-time central engine activity, either through internal shocks \citep{IKZ05, FW05, Zhang06} or other dissipation processes within the relativistic outflow \citep{Giannios06, Lazzati+11}. Proposed mechanisms include continued central engine activity \citep{Dai&Lu2002, Meszaros2002},
delayed reactivation of the central engine
\citep{MacFadyen+2001, King+2005, Margutti+2011, Wang&Dai2013, Dall_Osso+2017, Gibson+2018}, 
and delayed magnetic dissipation during ejecta deceleration \citep{Giannios06}. 

Alternative models rely on external outflow characteristics, such as refreshed shocks \citep{RM98, Laskar+15},
patchy shells \citep[e.g.,][]{RM98, ZM02},
or density fluctuations in the circumburst medium \citep{Wang&Loeb2000, Dai&Lu2002}. 
Other scenarios include slightly misaligned lines of sight to structured jets \citep{DBD+22}, or episodic accretion onto a black hole, as described in accretion disk models \citep{King+2005, Perna+2006, Proga&Zhang2006}.
In the later framework, instabilities or fragmentations within the accretion disk can lead to intermittent accretion episodes, producing the observed X-ray flares.

Several studies have suggested that the temporal structure of X-ray flares may reflect intermittent or fragmented accretion processes operating over a broad range of timescales. In particular, episodic accretion models predict asymmetric emission episodes with rapid rise phases followed by slower decays, naturally producing variability patterns similar to those observed in both prompt pulses and X-ray flares \citep{NBK05, King+2005, Perna+2006, Proga&Zhang2006, Kumar+2008a}. The stochastic waiting-time behavior observed from prompt pulses to X-ray flares \citep{Guidorzi+15} further suggests that GRB central engines may operate through intermittent episodes of activity, possibly associated with episodic accretion, rather than through a single continuous emission phase. However, \citet{Guidorzi+15} considered the overall waiting-time distribution without distinguishing X-ray flares according to their temporal morphology. Whether different flare populations occupy distinct regimes within this common stochastic process therefore remains unexplored.

The observed diversity of flares - early and late, wide and narrow, single or multiple 
\citep{Chincarini+10, Margutti+2011} challenges theoretical models. For example, it is unknown whether this diversity reflects a continuous temporal evolution or physically distinct flare populations. Considering these populations separately is crucial, as some theoretical models can only account for one type of flares. Internal dissipation scenarios, for example, require flares to exhibit very short variability timescales ($\delta t/t \ll 0.1$), which strongly favors early flares \citep[e.g.,][]{KZ15, Nakar&Piran2002, IKZ05, Nakar&Granot2007}. In contrast, external shock scenarios are constrained by the minimum variability timescale expected from adiabatic expansion in a wind environment: \citet{LP07} showed that $w/t_{\rm pk}$ values below $\sim0.83$ are difficult to achieve in this case. Consequently, late flares are more 
difficult to explain within prompt-like internal dissipation scenarios characterized by short variability timescales, whereas early flares are more naturally interpreted by such models. Indeed, some models, such as the one proposed by \citet{DBD+22}, can reproduce only the early ($< 1000$ s) flares, highlighting the importance of distinguishing between these populations when evaluating flare origins. 

Our previous studies provide further motivation for investigating the temporal diversity of X-ray flares. In \citet{DB+25}, we analyzed 89 GRBs with measured redshift,\footnote{\url{https://www.mpe.mpg.de/~jcg/grbgen.html}} 61 of which exhibit X-ray flares, and characterized their temporal properties through a simultaneous decomposition of the flare and underlying afterglow emission. In that work, we showed that the flare properties are largely independent of the presence of an X-ray plateau. We subsequently identified additional temporal features, including a bimodal distribution of the flare asymmetry parameter $t_{\rm rise}/t_{\rm decay}$, together with correlations among the characteristic flare timescales, such as flare width and peak time \citep{DB+26}. These results indicate a complex temporal behavior extending from early, rapidly varying flares to broader events occurring at later times and motivate the question of whether this diversity can be described by a single continuously evolving flare population or instead reveals statistically distinct temporal regimes. In the present work, we address this question by jointly examining the flare occurrence times and characteristic variability timescales, and investigate how the resulting populations differ in their temporal and energetic properties and what these differences imply for the physical origin of GRB X-ray flares. 

This paper is organized as follows. In Section~\ref{sec:flare_sampling}, we briefly summarize the GRB sample and flare characterization adopted from our previous work \citep{DB+25}. In Section~\ref{sec:flare_properties}, we investigate the temporal and energetic properties of the flares and characterize the early--narrow and late--wide populations. In Section~\ref{sec:discussion}, we discuss the properties of the two populations in the context of previous flare and prompt-emission studies and proposed physical models for X-ray flare production. In Section~\ref{sec:summary}, we summarize our main results and conclusions. Finally, in Appendix~\ref{sec:robustness_classification}, we assess the robustness of the flare classification and temporal properties against possible observational and methodological effects.
Throughout this paper, we adopt a flat $\Lambda$CDM cosmology with $\Omega_{\rm m}=0.286$ and $H_0=70~{\rm km~s^{-1}~Mpc^{-1}}$ \citep{Hinshaw+09}.

\section{Sample and Flare Characterization}
\label{sec:flare_sampling}

Our analysis builds directly on the GRB sample and flare-fitting methodology presented in \citet{DB+25}. From an initial sample of 100 \textit{Swift}-BAT GRBs with known redshift selected in reverse chronological order from December 2022, 89 events were selected after light-curve fitting (the other 11 were excluded as the flare width was wider than available lightcurve, that prevented complete unambiguous fit to the entire lightcurve; see \citet{DB+25} for full details). Among these, 61 GRBs exhibit at least one X-ray flare and form the focus of this study.

Flares were modeled with the asymmetric Norris profile \citep{NBK05}, following the full formulation and parameterization described and presented in \citet{DB+25} and summarized in \citet{DB+26}. This modeling allowed us to measure the flare peak time $t_{\rm pk}$, the flare width $w$, asymmetry, and the flare isotropic energy $E_{\rm iso,f}$. The width was defined in log-space based on half-maximum limits, and the ratio $w/t_{\rm pk}$ provides a robust description of flare duration and variability. These measurements constitute the basis for our investigation of flare populations.
The present work builds upon these published measurements to investigate the statistical properties of the flare population and assess whether distinct subpopulations can be identified and physically interpreted.

The flare parameters and their uncertainties used throughout this work, including the flare peak time ($t_{\rm pk}$), width ($w$), rise time ($t_{\rm rise}$), decay time ($t_{\rm decay}$), asymmetry, isotropic energy ($E_{\rm iso,f}$), and redshift are taken from the machine-readable tables presented in \citet{DB+25, DB+26}. The present work builds upon these published measurements to investigate the statistical properties and physical interpretation of the early--narrow and late--wide flare populations.

\section{Flare properties}
\label{sec:flare_properties}
\subsection{Properties of single and double flares}
\begin{figure}%[ht!]
\centering
\includegraphics[width=0.45\linewidth]{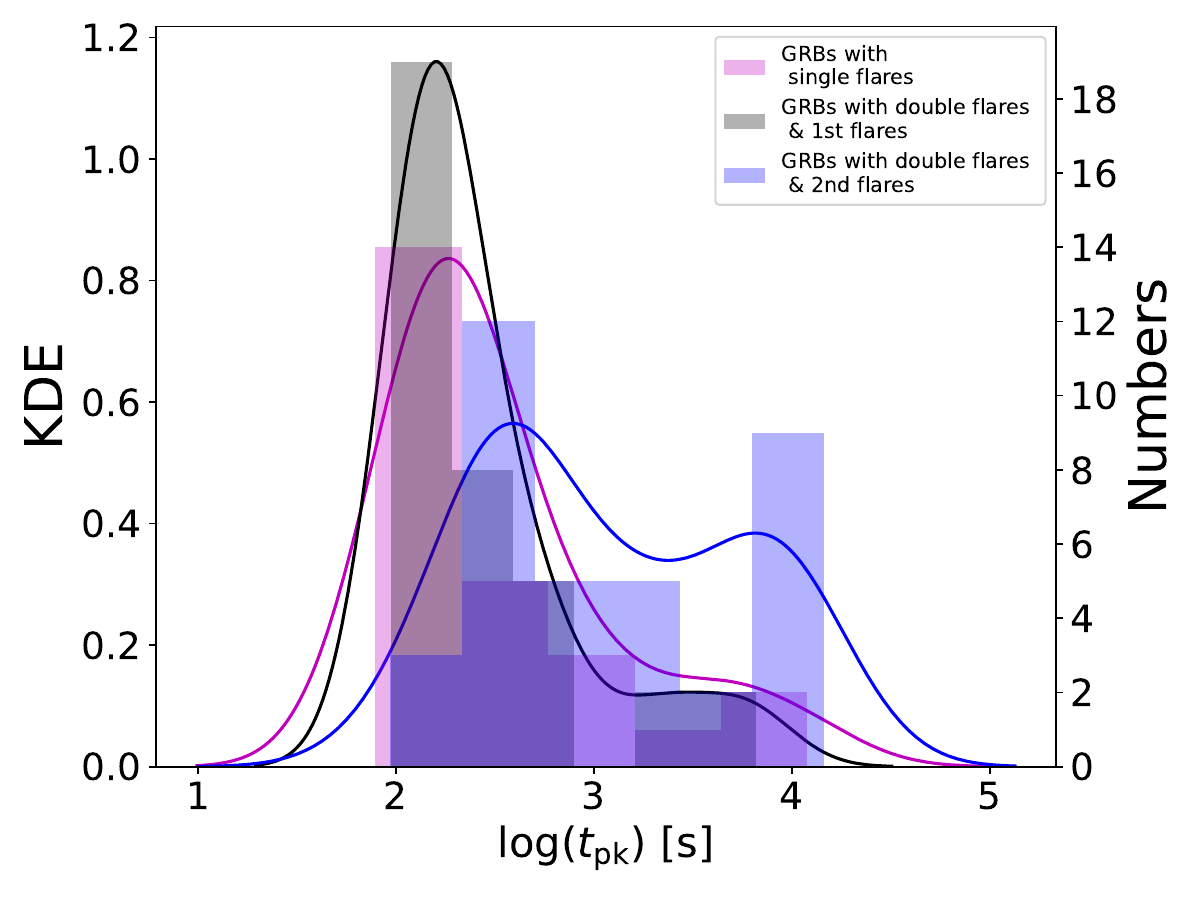}
\includegraphics[width=0.45\linewidth]{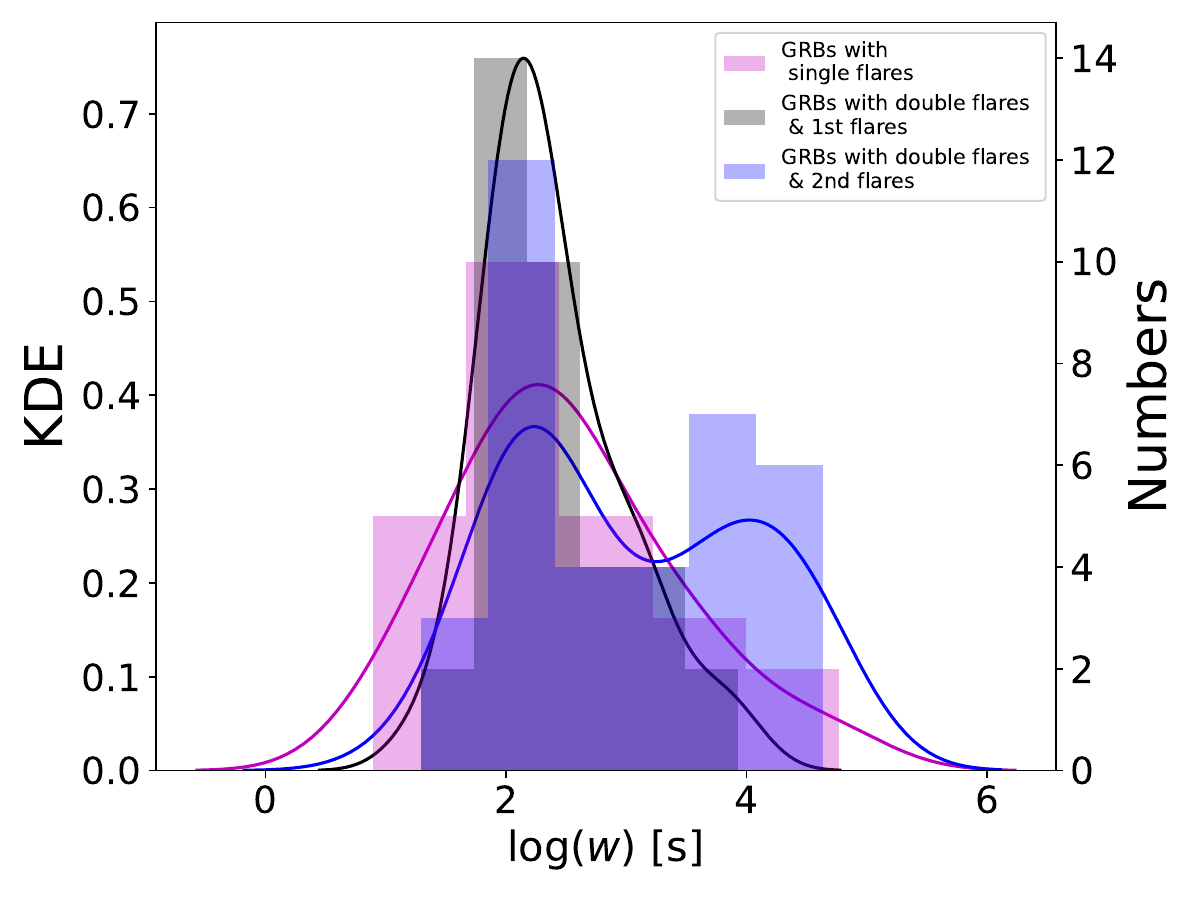}
\includegraphics[width=0.45\linewidth]{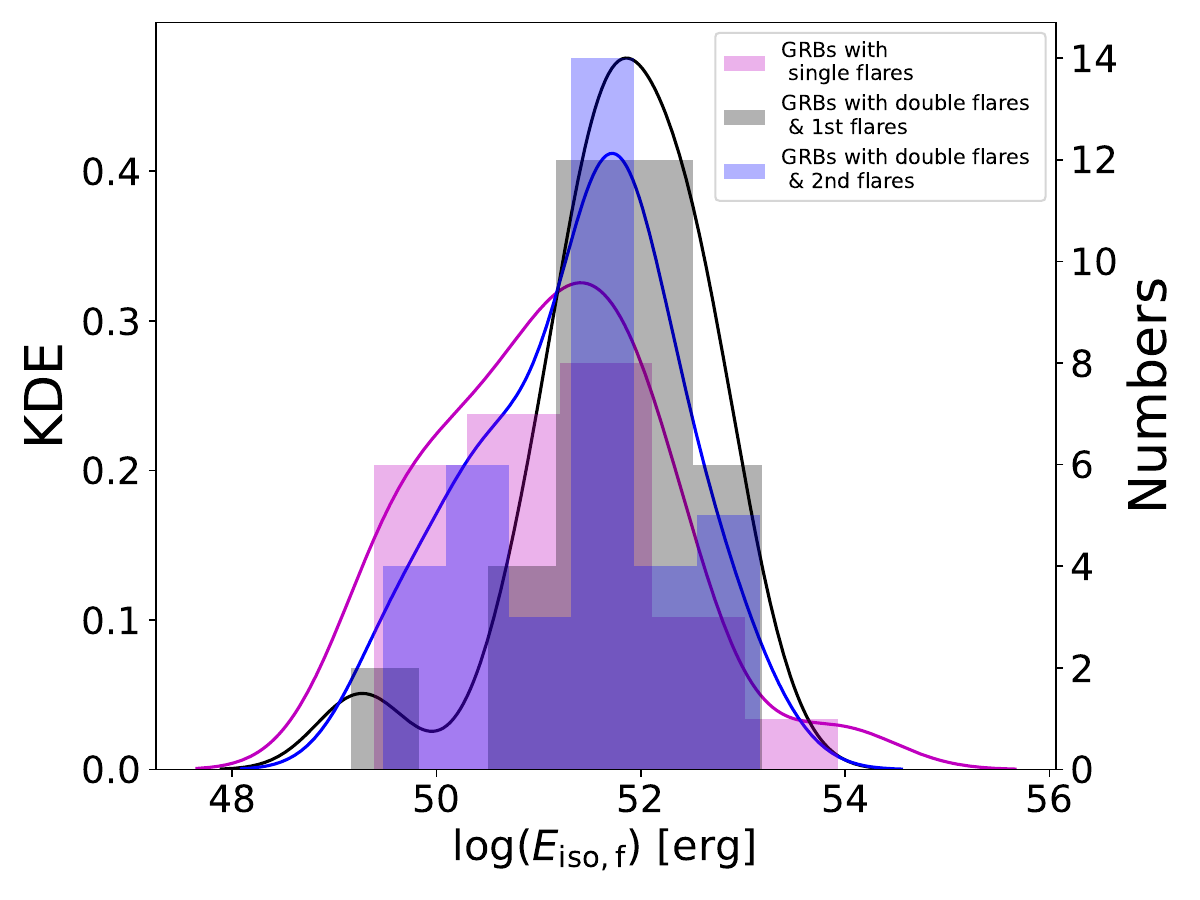}
\caption{Distributions of the flare peak time, $t_{\rm pk}$ (top left), the flare width, $w$ (top right), and the flare isotropic energy, $E_{\rm iso,f}$ (bottom), for GRBs with single and double flares. The flares in double-flare GRBs are further separated into first and second flare episodes. In each panel, the right-hand ordinate shows the number of flares in each bin, while the left-hand ordinate shows the kernel density estimate (KDE) drawn by the purple, black and blue solid lines respectively for each subsample.}
\label{fig:hist_peak-time_width_Eiso-flare}
\end{figure}

In our previous studies \citep{DB+25,DB+26}, we considered the distributions of flare peak time and width, in GRBs with and without a plateau phase. We found that both GRB samples showed extended tails toward large values of $t_{\rm pk}$ and $w$. Here, we first investigate whether these tails can be explained by flare order, in particular whether the large values of $t_{\rm pk}$ and $w$ are associated exclusively with second flare episodes. We combine the data of all GRBs (both with and without a plateau). 
In Figure~\ref{fig:hist_peak-time_width_Eiso-flare} (top left and top right), we compare the flare peak-time ($t_{\rm pk}$) and flare-width ($w$) distributions for GRBs with a single flare and for the first and second flares of GRBs exhibiting two flare episodes. The distributions show that the tail toward larger peak times and widths is not composed solely of second flares, although second flares contribute preferentially to this region. This behavior is particularly pronounced in the flare peak-time distribution.

We further compare the statistical properties of the flare peak time ($t_{\rm pk}$), flare width ($w$), and flare isotropic energy ($E_{\rm iso,f}$) for single, first, and second flares in Table~\ref{tab:flare_statistics}. The peak-time distributions of single and first flares are statistically consistent ($p=0.49$), whereas second flares occur systematically later than both first ($p=2\times10^{-5}$) and single flares ($p=4\times10^{-4}$). In contrast, the width distributions do not differ significantly among the three groups. The isotropic-energy distributions of first and second flares are also statistically consistent, whereas the single- and first-flare distributions show a statistically significant difference ($p=0.01$).

The bottom panel of Figure~\ref{fig:hist_peak-time_width_Eiso-flare} shows the corresponding distributions of flare isotropic energy. In contrast to the tails toward larger values of $t_{\rm pk}$ and $w$, the energy distribution shows a tail extending toward lower values of $E_{\rm iso,f}$, with later second flares tending to have lower average isotropic energies than first flares. However, the isotropic-energy distributions of the first and second flares are statistically consistent ($p=0.49$), indicating that this tendency is not statistically significant. The relation between flare energetics and temporal properties is investigated further in Section~\ref{sec:flare energy}.

Thus, although second flares occur systematically later and contribute preferentially to the tail toward large $t_{\rm pk}$ and $w$, flare order alone cannot explain the observed distributions. This motivates examining the joint temporal properties of the flares rather than classifying them solely according to their order of occurrence.
%%%%%%%%%%%%%%%%%%%%%%%%%%%%%%%%%%%%%%%%%%%%%%%%%%%%%%%%%%
%%%%%%%% Figure: $t_{\rm pk}$, $w$, and $E_{\rm iso,f}$
%%%%%%%%%%%%%%%%%%%%%%%%%%%%%%%%%%%%%%%%%%%%%%%%%%%%%%%%%%

%%%%%%%%%%%%%%%%%%%%%%%%%%%%%%%%%%%%%%%%%%%%%%%%%%%%%%%%%%
%%%%%%%% 
%%%%%%%%%%%%%%%%%%%%%%%%%%%%%%%%%%%%%%%%%%%%%%%%%%%%%%%%%%

%%%%%%%%%%%%%%%%%%%%%%%%%%%%%%%%%%%%%%%%%%%%%%%%%%%%%%%%%%
%%%%%%%% Table: Statistics
%%%%%%%%%%%%%%%%%%%%%%%%%%%%%%%%%%%%%%%%%%%%%%%%%%%%%%%%%%
\begin{table}[H]
\centering
\caption{Statistics of the flare properties for GRBs with a single flare and for the first and second flares of GRBs exhibiting two flare episodes. The average values are reported with their standard errors of the mean (SEM), while $\sigma$ denotes the standard deviation of the corresponding distribution. The three K--S test columns give the comparisons between single and first flares, first and second flares, and single and second flares, respectively. The number of second flares is identical to that of first flares and is therefore not repeated.}
\begin{tabular}{lc c c c c cc c c c c }
\toprule
& \multicolumn{3}{c}{Single flares} 
& \multicolumn{3}{c}{1st flares} 
& \multicolumn{2}{c}{2nd flares}  \\
\cmidrule(lr){2-4} \cmidrule(lr){5-7} \cmidrule(lr){8-9} 
\textbf{Parameter} 
& $N_{\rm flares}$ & Avg $\pm$ SEM ($\sigma$)& $\begin{array}{c}\text{K-S}\\(D,p)\end{array}$ 
& $N_{\rm flares}$ &  Avg $\pm$ SEM ($\sigma$) & $\begin{array}{c}\text{K-S}\\(D,p)\end{array}$ 
&  Avg $\pm$ SEM ($\sigma$)& $\begin{array}{c}\text{K-S}\\(D,p)\end{array}$ 
 \\
\midrule
$\left< \log_{10} (t_{\rm pk}) \right>$ 
& 25 & 2.52$\pm$0.11 (0.57) & (0.20, 0.49) 
& 36 & 2.42$\pm$0.08 (0.47) & (0.72, 0.00002) 
& 3.08$\pm$0.11 (0.66) & (0.52, 0.0004)  \\
$\left< \log_{10} (w) \right>$ 
& 25 & 2.49$\pm$0.19 (0.93) & (0.18, 0.66) 
& 36 & 2.39$\pm$0.10 (0.58) & (0.31, 0.26) 
& 3.03$\pm$0.17 (1.01) & (0.28, 0.16)  \\
$\left< \log_{10} (E_{\rm iso}) \right>$ 
& 25 & 51.10$\pm$0.22 (1.10) & (0.41, 0.01) 
& 36 & 51.78$\pm$0.14 (0.86) & (0.25, 0.49) 
& 51.37$\pm$0.16 (0.95) & (0.21, 0.45)  \\
\bottomrule
\end{tabular}
\label{tab:flare_statistics}
\end{table}

%%%%%%%%%%%%%%%%%%%%%%%%%%%%%%%%%%%%%%%%%%%%%%%%%%%%%%%%%%
%%%%%%%% Figure: correlation between $t_{\rm pk}$ and $w$
%%%%%%%%%%%%%%%%%%%%%%%%%%%%%%%%%%%%%%%%%%%%%%%%%%%%%%%%%%

\begin{figure}%[ht!]
\centering
\includegraphics[width=\linewidth]{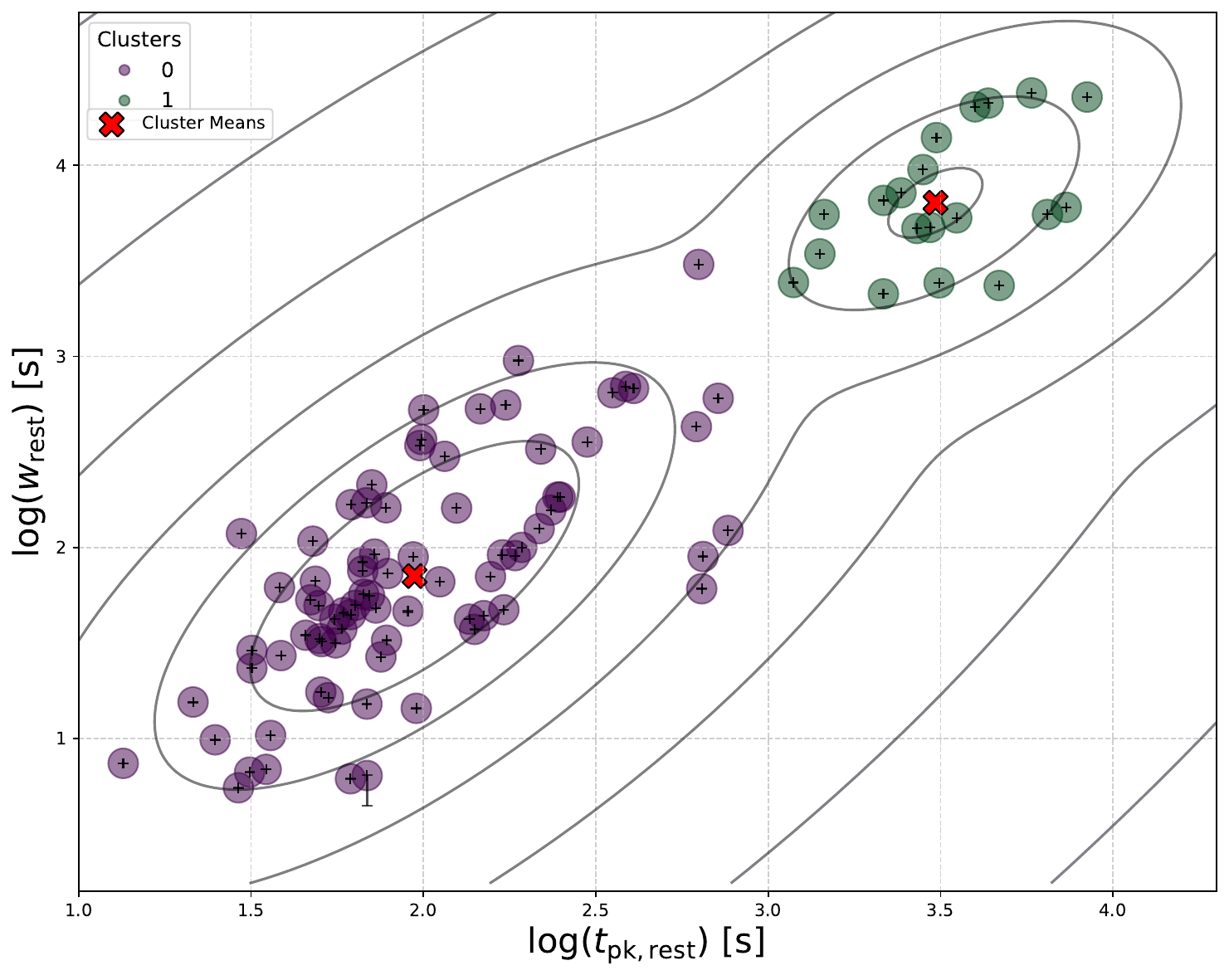}
\caption{Relation between the rest-frame flare peak time ($t_{\rm pk}$) and flare width ($w$), with a two-component Gaussian Mixture Model (GMM) applied to the data. The data are shown in logarithmic space, with error bars corresponding to the $1\sigma$ uncertainties. The purple and green points represent the early--narrow (Cluster~0) and late--wide (Cluster~1) flare populations, respectively. The contours show the probability density of the two Gaussian components, while the ``$\times$'' symbols indicate the centers of the corresponding clusters.}
\label{fig:scatter_tpk_width}
\end{figure}

%%%%%%%%%%%%%%%%%%%%%%%%%%%%%%%%%%%%%%%%%%%%%%%%%%%%%%%%%%
%%%%%%%%
%%%%%%%%%%%%%%%%%%%%%%%%%%%%%%%%%%%%%%%%%%%%%%%%%%%%%%%%%%

\subsection{Identification of Distinct Flare Populations}
\label{sec:narrow_wide_flares}

To investigate whether the extended distributions toward large flare peak times and widths reflect statistically distinct components, we first applied a univariate Gaussian Mixture Model (GMM) separately to the observer-frame distributions of $\log(t_{\rm pk})$ and $\log(w)$ presented in our previous studies \citep{DB+25,DB+26}. The resulting fits revealed two prominent components in each distribution. We first carried the analysis separately for GRBs with and without a plateau. For $\log(t_{\rm pk})$, the peaks are centered at $2.31 ~(\sigma =0.25)$ and $3.60 ~(\sigma =0.36)$ for GRBs with a plateau, and at $2.41 ~(\sigma =0.26)$ and $3.63 ~(\sigma =0.41)$ for those without. Similarly, in the $\log(w)$ distributions, the peaks appear at $2.22 ~(\sigma =0.49)$ and $3.77 ~(\sigma =0.38)$ (with plateau) and at $2.25 ~(\sigma =0.49)$ and $4.52 ~(\sigma =0.18)$ (without plateau). These results support the interpretation that both early--narrow and late--wide flares coexist within each GRB population (with and without plateaus), with late--wide flares contributing predominantly to the high-value tails of the distributions.
Importantly, the peak positions and overall distributions are consistent between GRBs with and without a plateau phase, indicating no significant dependence on the presence of a plateau. Therefore, we do not treat the two samples separately and do not further distinguish between plateau and non-plateau GRBs when studying early--narrow and late--wide flares.

To further explore whether these flare types form statistically distinct clusters in a multidimensional parameter space, we applied a multivariate GMM to the two-dimensional distribution (correlation) of $\log(t_{\rm pk})$ and $\log(w)$ in the rest frame. Performing the clustering in the rest frame, using the GRB redshifts listed in Table B1 of \citet{DB+26}, reduces the effect of cosmological time dilation and produces a clearer separation between the early--narrow and late--wide populations, particularly in the flare-width distribution. We therefore adopt the rest-frame classification throughout the remainder of this work. The corresponding observer-frame analysis is presented and compared in Appendix~\ref{GMM_tpk_w_observer_frame}.

The optimal number of components was assessed using the Bayesian Information Criterion (BIC) and the Akaike Information Criterion (AIC). The BIC reaches its minimum for a two-component model, while the AIC does not provide a significant improvement when increasing the number of components from two to three. We therefore adopt a two-component GMM, which provides an adequate description of the joint distribution, as shown in Figure~\ref{fig:scatter_tpk_width}. Although a three-component GMM was also explored, it primarily subdivides the broad lower distribution into two overlapping groups rather than identifying a clearly distinct third population.
~

From the multivariate GMM parameters, we identify two distinct clusters:
\begin{itemize}
\item \textbf{Cluster 0}, the dominant population (80\%), is centered around $\log_{10}(t_{\rm pk}) = 1.97$ and $\log_{10}(w) = 1.85$, corresponding to earlier and narrower flares. This component exhibits a strong positive correlation ($\rho \simeq 0.68$) between flare peak time and width.
\item \textbf{Cluster 1} represents approximately 20\% of the sample and has mean values of $\log_{10}(t_{\rm pk}) = 3.49$ and $\log_{10}(w) = 3.80$, indicating a population of wide, late-time flares. This component also shows a positive correlation between the two parameters ($\rho \simeq 0.54$), suggesting that wider flares tend to occur at later times.
\end{itemize}
These findings reinforce the presence of two physically distinct flare populations, with the late-wide-flare population peaking systematically at later times than the early-narrow-flare population.

We further find that the distribution of flare order differs between the two flare populations. Among the 61 flares identified as either single or first flares in their light curves, 55 (90\%) belong to the early--narrow population (Cluster~0) and only 6 (10\%) to the late--wide population (Cluster~1). In contrast, among the 36 second flares, 23 (64\%) belong to the early--narrow population and 13 (36\%) to the late--wide population. Thus, although approximately 70\% of the late--wide population consists of second flare episodes (13 out of 19; 68\%), second flares are distributed across both populations, with the majority (23 out of 36; 64\%) still belonging to the early--narrow population. These results indicate that, while flare order is associated with cluster membership, it cannot by itself explain the distinction between the two populations.

When comparing the results of the univariate (distribution) and multivariate GMM (correlation) analyses, we find that the characteristic scales of $\log_{10}(t_{\rm pk})$ are highly consistent between the two methods, with both identifying peaks and clusters at similar values. A similar consistency is observed for $\log_{10}(w)$, where both approaches reveal two distinct regimes corresponding to early--narrow and late--wide flares. 
However, while the univariate analysis identifies bimodality in each parameter separately, the multivariate GMM reveals coherent clusters in the joint parameter space, thereby providing a more physically meaningful classification of flare populations. We therefore conclude that the multivariate approach offers a more robust framework for distinguishing between early--narrow and late--wide flares.
This classification is also broadly consistent with previous approaches based on a temporal cut at $t_{\rm pk} < 1000$ s \citep{Chincarini+10, Bernardini+11}, as illustrated in Figure~\ref{fig:scatter_tpk_width}.
 
\subsubsection{Relations between flare energetics and temporal properties}
\label{sec:flare energy}

\begin{figure}%[ht!]
\centering
\includegraphics[width=0.3\linewidth]{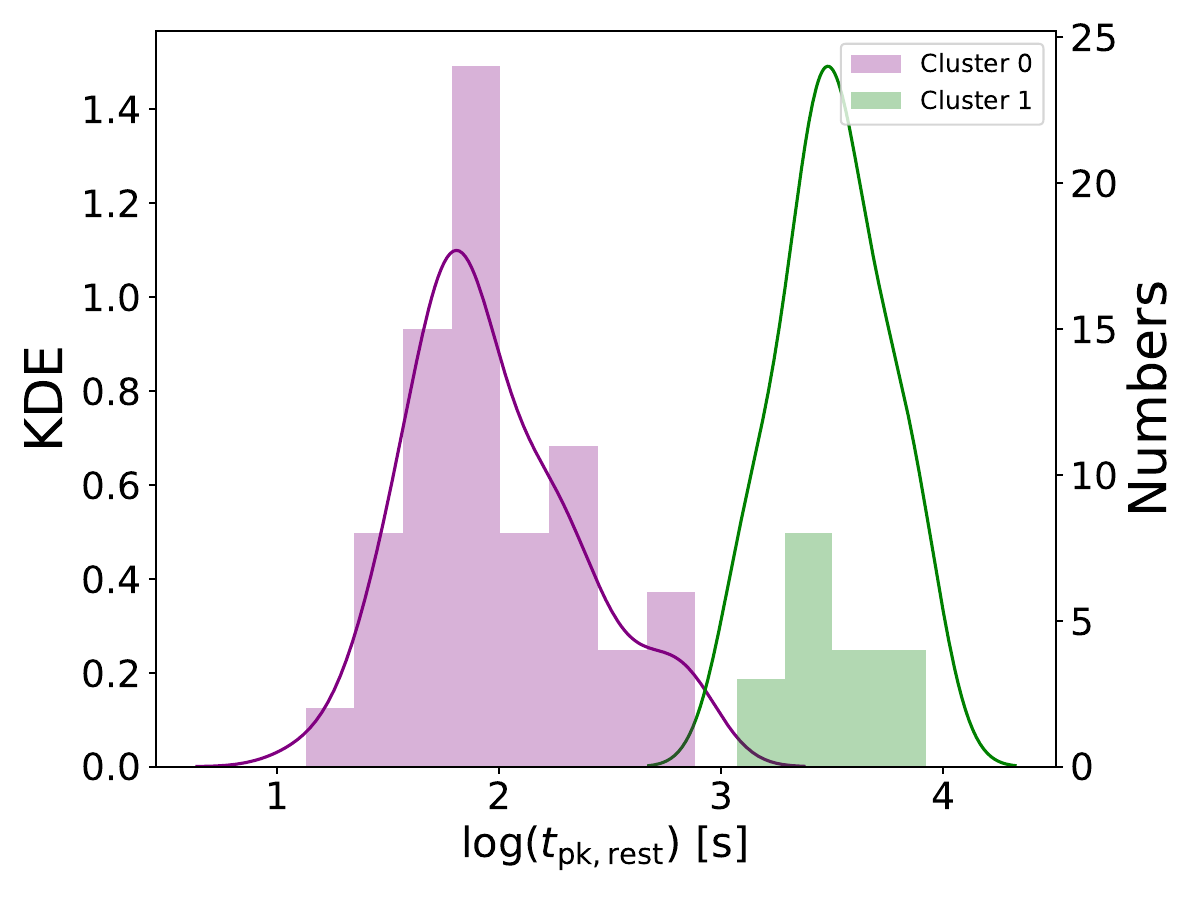}
\includegraphics[width=0.3\linewidth]{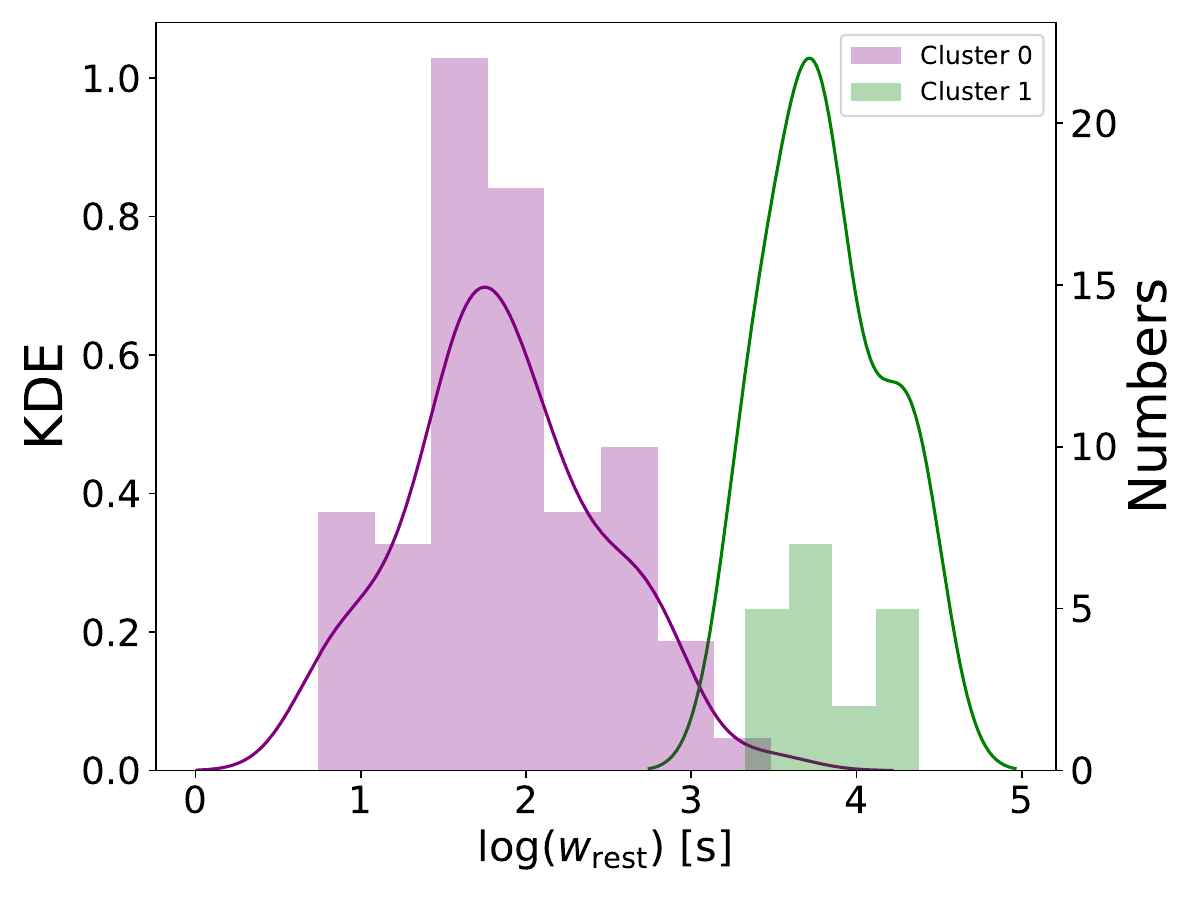}
\includegraphics[width=0.3\linewidth]{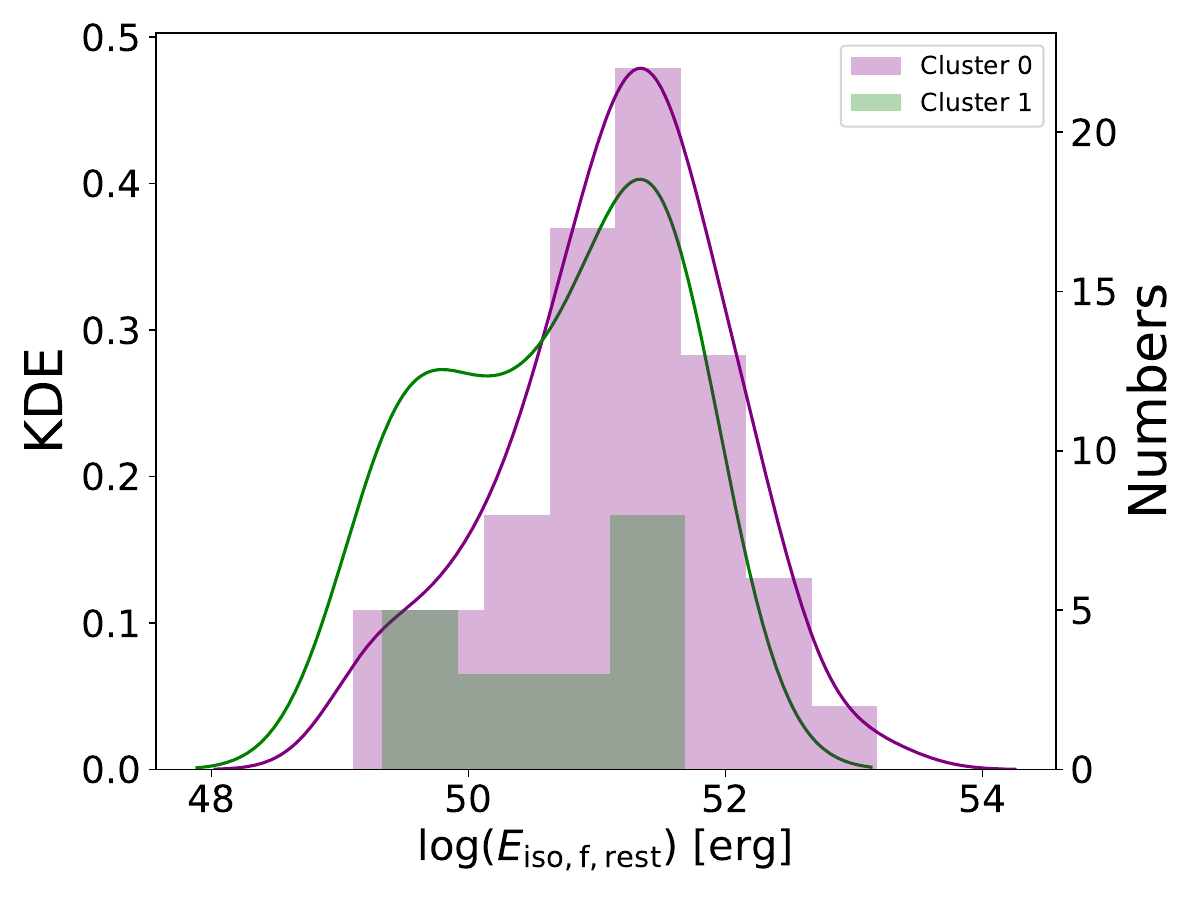}
\caption{Distributions of the rest-frame flare peak time, $t_{\rm pk, rest}$ (left), flare width, $w_{\rm rest}$ (middle), and flare isotropic energy, $E_{\rm iso,f,rest}$ (right), for the early--narrow (Cluster~0) and late--wide (Cluster~1) flare populations identified by the Gaussian Mixture Model. Cluster~0 (purple) contains 78 flares ($\sim80\%$ of the sample), while Cluster~1 (green) contains 19 flares ($\sim20\%$). In each panel, the right-hand ordinate shows the number of flares in each bin, while the left-hand ordinate shows the corresponding kernel density estimate (KDE).}
\label{fig:hist_peak-time_width_Eiso-components}
\end{figure}

Figure~\ref{fig:hist_peak-time_width_Eiso-components} shows the distributions
of the flare peak times $t_{\rm pk,rest}$, the flare widths $w_{\rm rest}$, and the flare isotropic energies $E_{\rm iso,f,rest}$ in the rest frame for both early--narrow and late--wide flares. The statistical results for these parameters, including the number of GRBs, average values, and Kolmogorov–Smirnov (K-S) test outcomes, are summarized in Tables~\ref{tab:narrow-wide_statistics}.

The separation between early--narrow and late--wide flares is evident in the $t_{\rm pk, rest}$ and $w_{\rm rest}$ distributions, as the GMM applied to these parameters defines the flare types. In contrast, the separation in the $E_{\rm iso,f,rest}$ distribution is less pronounced. Although both subsamples show a low-energy tail, it is more dominant in the distributions of GRBs with late--wide flares. Moreover, GRBs with late--wide flares tend to show a lower average isotropic energy ($\left < \rm log_{10}(E_{\rm iso,f, rest}) \right > = 50.66\pm0.20, \sigma = 0.87 $ vs $\left < \rm log_{10}(E_{\rm iso,f, rest}) \right > = 51.13\pm0.10, \sigma = 0.85$, however, the K--S test yields $D = 0.26$ and $p = 0.22$. Since $p > 0.05$, we cannot reject the null hypothesis at the $5\%$ significance level, suggesting that the two subsamples are consistent with being drawn from the same population) compared to the GRBs with early--narrow flares.

%%%%%%%%%%%%%%%%%%%%%%%%%%%%%%%%%%%%%%%%%%%%%%%%%%%%%%%%%%
%%%%%%%% Figure: Distributions of $t_{\rm pk}$, $w$, and $E_{\rm iso,f}$
%%%%%%%%%%%%%%%%%%%%%%%%%%%%%%%%%%%%%%%%%%%%%%%%%%%%%%%%%%

%%%%%%%%%%%%%%%%%%%%%%%%%%%%%%%%%%%%%%%%%%%%%%%%%%%%%%%%%%
%%%%%%%%
%%%%%%%%%%%%%%%%%%%%%%%%%%%%%%%%%%%%%%%%%%%%%%%%%%%%%%%%%%

%%%%%%%%%%%%%%%%%%%%%%%%%%%%%%%%%%%%%%%%%%%%%%%%%%%%%%%%%%
%%%%%%%% Table : Statistics of the narrow-wide flare properties :
               % N_flares and Avg values
%%%%%%%%%%%%%%%%%%%%%%%%%%%%%%%%%%%%%%%%%%%%%%%%%%%%%%%%%%
\begin{table}[H]
\centering
\caption{Statistics of the early--narrow and late--wide flare properties: Number of flares, average values with their standard errors of the mean (SEM) and standard deviations ($\sigma$), and Kolmogorov--Smirnov (K--S) test results, reporting the K--S statistic $D$ and the associated probability $p$ for early--narrow (Cluster 0) and late--wide (Cluster 1) flare properties. The parameters we show are 
(i) the flare peak time, $t_{\rm pk, rest}$, (ii) the flare width $w_{\rm rest}$ (iii) the flare isotropic energy $E_{\rm iso,f, rest}$ respectively. Average values indicate that early--narrow and late--wide flares differ significantly in both $t_{\rm pk,rest}$ and $w_{\rm rest}$, consistent with the two-component GMM classification applied to the joint $\log(t_{\rm pk, rest})$--$\log(w_{\rm rest})$ parameter space. In contrast, although late--wide flares exhibit a lower average isotropic energy than early--narrow flares, the K--S test ($p=0.22$) indicates that the difference in $E_{\rm iso,f}$ is not statistically significant.}
\begin{tabular}{l|cc|cc|cc}
\toprule
 & \multicolumn{2}{c|}{Cluster 0: early--narrow Flares} 
 & \multicolumn{2}{c}{Cluster 1: late--wide Flares} 
 & \multicolumn{2}{c}{K--S test} \\
\cmidrule(lr){2-3} \cmidrule(lr){4-5} \cmidrule(lr){6-7} 
\textbf{Parameter}
 & $N_{\rm flares}$ & Avg $\pm$ SEM ($\sigma$)
 & $N_{\rm flares}$ & Avg $\pm$ SEM ($\sigma$)
 &  D & p \\
\midrule
$ \log_{10} (t_{\rm pk, rest}) $  & 78 & 1.98$\pm$0.04 (0.39) &  19 & 3.50$\pm$0.05 (0.24) & 1.0 & $2.86\times10^{-20}$  \\
$ \log_{10} (w_{\rm rest}) $    & 78 & 1.86$\pm$0.07 (0.59) & 19 & 3.82$\pm$0.08 (0.35) &0.99 & $ 4.73\times 10^{-19}$ \\
$ \log_{10} (E_{\rm iso, f, rest}) $  & 78 & 51.13$\pm$0.10 (0.85)& 19 & 50.66$\pm$0.20 (0.87) & 0.26& 0.22  \\
\bottomrule
\end{tabular}
\label{tab:narrow-wide_statistics}
\vspace{5mm}
%\parbox{\textwidth}{\small}
\end{table}

To further investigate whether this apparent difference reflects a physical trend with time or flare width, we performed correlation analyses between $E_{\rm iso,f,rest}$ and both $t_{\rm pk,rest}$ and $w_{\rm rest}$ in the rest frame without separating the first and second flares.
We find that the Spearman rank correlation coefficient between $t_{\rm pk,rest}$ and $E_{\rm iso,f,rest}$ is $r=-0.34$ ($p=0.002$) for the early--narrow flares and $r=0.24$ ($p=0.30$) for the late--wide flares, indicating a weak but statistically significant negative correlation for the early--narrow flares and no statistically significant correlation for the late--wide flares.
Similarly, the Spearman rank correlation coefficient between the flare width ($w_{\rm rest}$) and the isotropic flare energy ($E_{\rm iso,f,rest}$) is $r=-0.04$ ($p=0.71$) for the early--narrow flares and $r=0.17$ ($p=0.50$) for the late--wide flares, indicating no statistically significant correlation between flare width and isotropic energy in either flare population.

\begin{figure}%[ht!]
\centering
\includegraphics[width=0.45\linewidth]{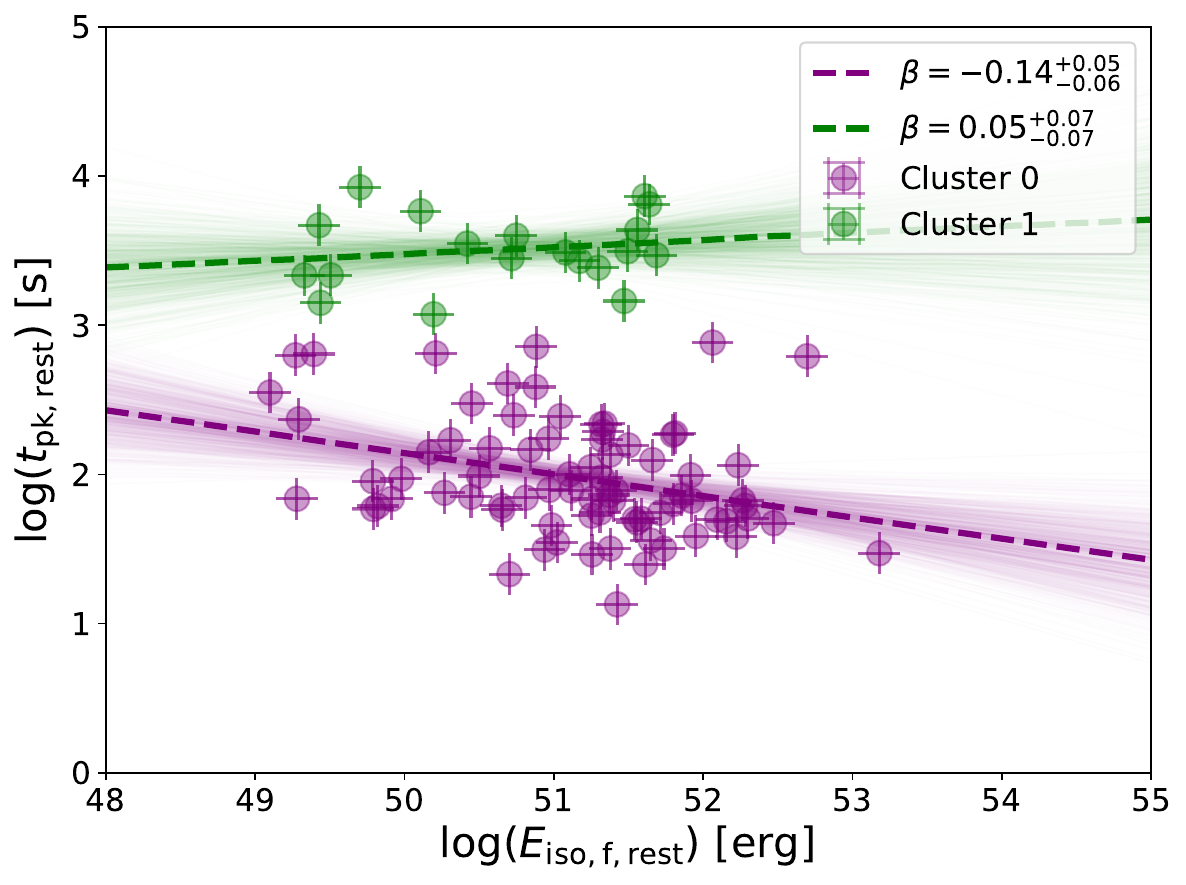}
\includegraphics[width=0.45\linewidth]{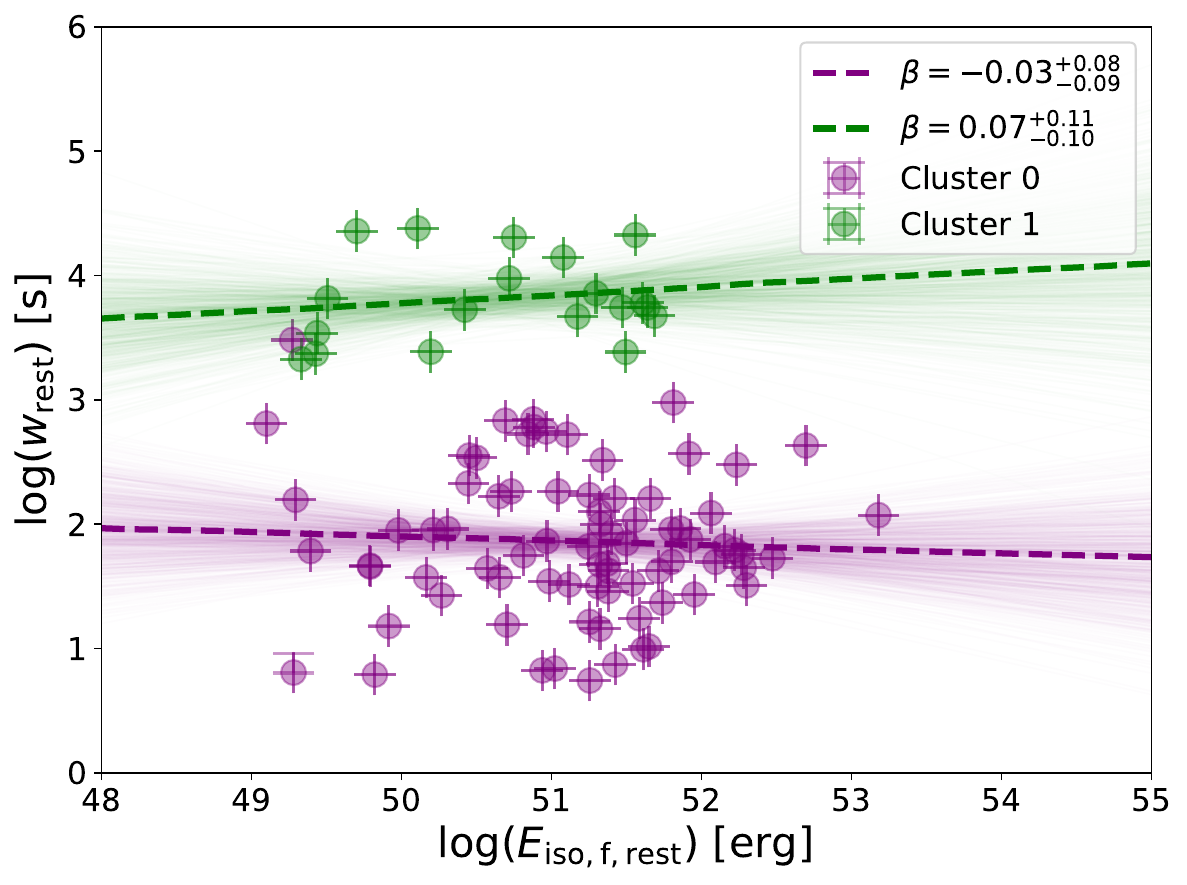}
\caption{Relation between the rest-frame flare peak time, $t_{\rm pk,rest}$, and the flare isotropic energy, $E_{\rm iso,f,rest}$ (left), and between the rest-frame flare width, $w_{\rm rest}$, and $E_{\rm iso,f,rest}$ (right). The purple and green points represent the early--narrow (Cluster~0; 78 flares) and late--wide (Cluster~1; 19 flares) flare populations, respectively. The error bars correspond to the $1\sigma$ uncertainties. The dashed lines show the median Bayesian linear regressions obtained using the method of \citet{Kelly2007}, which accounts for measurement uncertainties in both variables as well as intrinsic scatter.
For the $t_{\rm pk,rest}$--$E_{\rm iso,f,rest}$ relation, we obtain $\beta=-0.14^{+0.05}_{-0.06}$ ($\alpha=9.3^{+2.8}_{-2.4}$) for the early--narrow flares and $\beta=0.05^{+0.07}_{-0.07}$ ($\alpha=1.1^{+3.5}_{-3.7}$) for the late--wide flares. Similarly, for the $w_{\rm rest}$--$E_{\rm iso,f,rest}$ relation, we find $\beta=-0.03^{+0.08}_{-0.09}$ ($\alpha=3.6^{+4.4}_{-3.9}$) and $\beta=0.07^{+0.11}_{-0.10}$ ($\alpha=0.5^{+5.1}_{-5.3}$) for the early--narrow and late--wide flare populations, respectively. In both panels, the inferred slopes are consistent with zero within the $1\sigma$ credible intervals, indicating that the isotropic flare energy is not strongly coupled to either the flare peak time or flare width.}
\label{fig:scatter_tpk_w_Eiso}
\end{figure}

We then performed a Bayesian linear regression in log--log space using the method of \citet{Kelly2007}, which accounts for measurement uncertainties in both variables as well as intrinsic scatter as seen in Figure \ref{fig:scatter_tpk_w_Eiso}. For the $t_{\rm pk,rest}$--$E_{\rm iso,f,rest}$ relation, the inferred slope for the early--narrow flare population is mildly negative ($\beta=-0.14^{+0.05}_{-0.06}$), consistent with the weak anti-correlation found from the Spearman rank analysis, whereas the slope for the late--wide flare population is consistent with zero within the $1\sigma$ credible interval. For the $w_{\rm rest}$--$E_{\rm iso,f,rest}$ relation, the inferred slopes for both flare populations are consistent with zero within the $1\sigma$ credible intervals, indicating no statistically significant dependence of flare energetics on flare width.

Taken together, the K--S tests, Spearman rank correlations, and Bayesian linear regression analyses indicate that the flare isotropic energy shows at most a very weak dependence on flare peak time and no significant dependence on flare width. In particular, early--narrow flares exhibit a mild anti-correlation between $t_{\rm pk,rest}$ and $E_{\rm iso,f,rest}$, whereas no statistically significant trends are found for late--wide flares. Overall, the temporal properties of the two flare populations do not appear to be strongly linked to their energetics.

%%%%%%%%%%%%%%%%%%%%%%%%%%%%%%%%%%%%%%%%%%%%%%%%%%%%%%%%%%
%%%%%%%% Figure: Eiso vs tpk and width correlations
%%%%%%%%%%%%%%%%%%%%%%%%%%%%%%%%%%%%%%%%%%%%%%%%%%%%%%%%%%
%%%%%%%%%%%%%%%%%%%%%%%%%%%%%%%%%%%%%%%%%%%%%%%%%%%%%%%%%%
%%%%%%%%
%%%%%%%%%%%%%%%%%%%%%%%%%%%%%%%%%%%%%%%%%%%%%%%%%%%%%%%%%%

\begin{figure}%[ht!]
\centering
\includegraphics[width=0.45\linewidth]{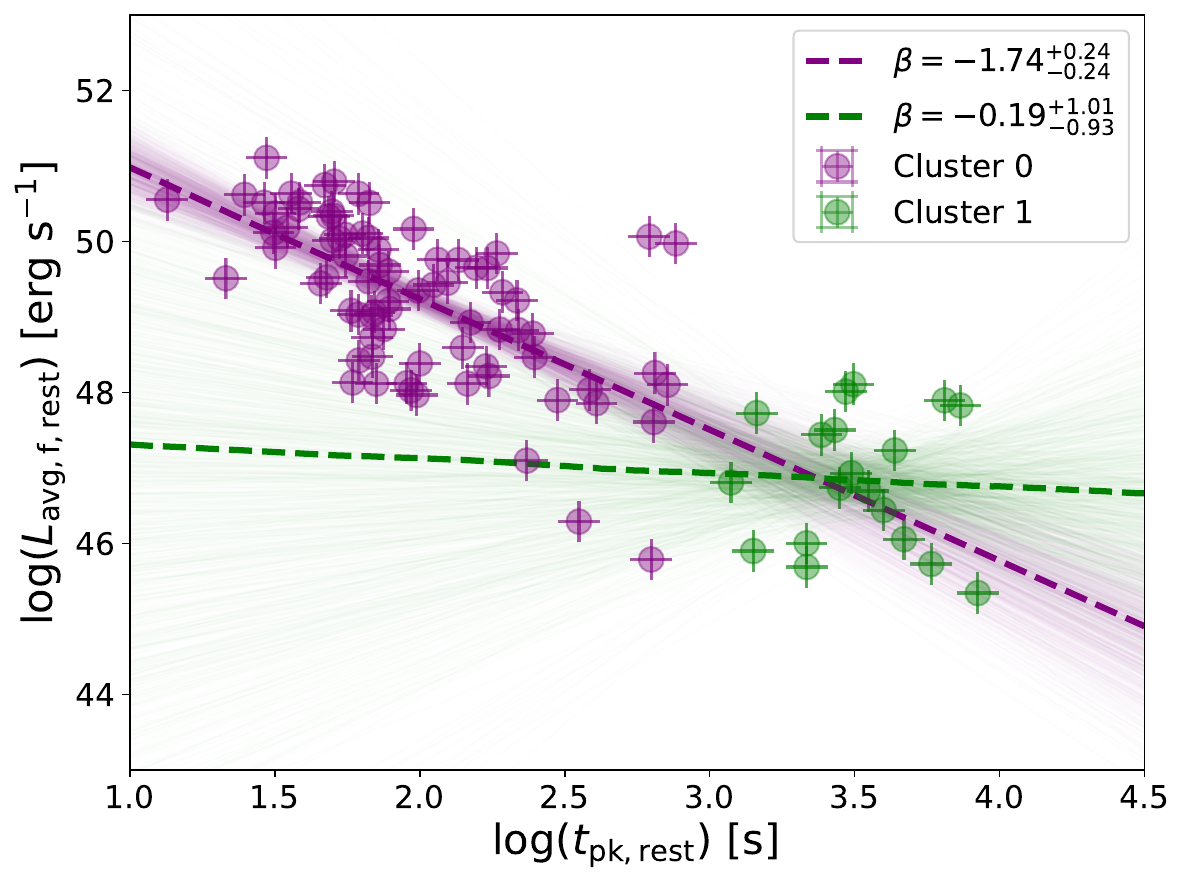}
\includegraphics[width=0.45\linewidth]{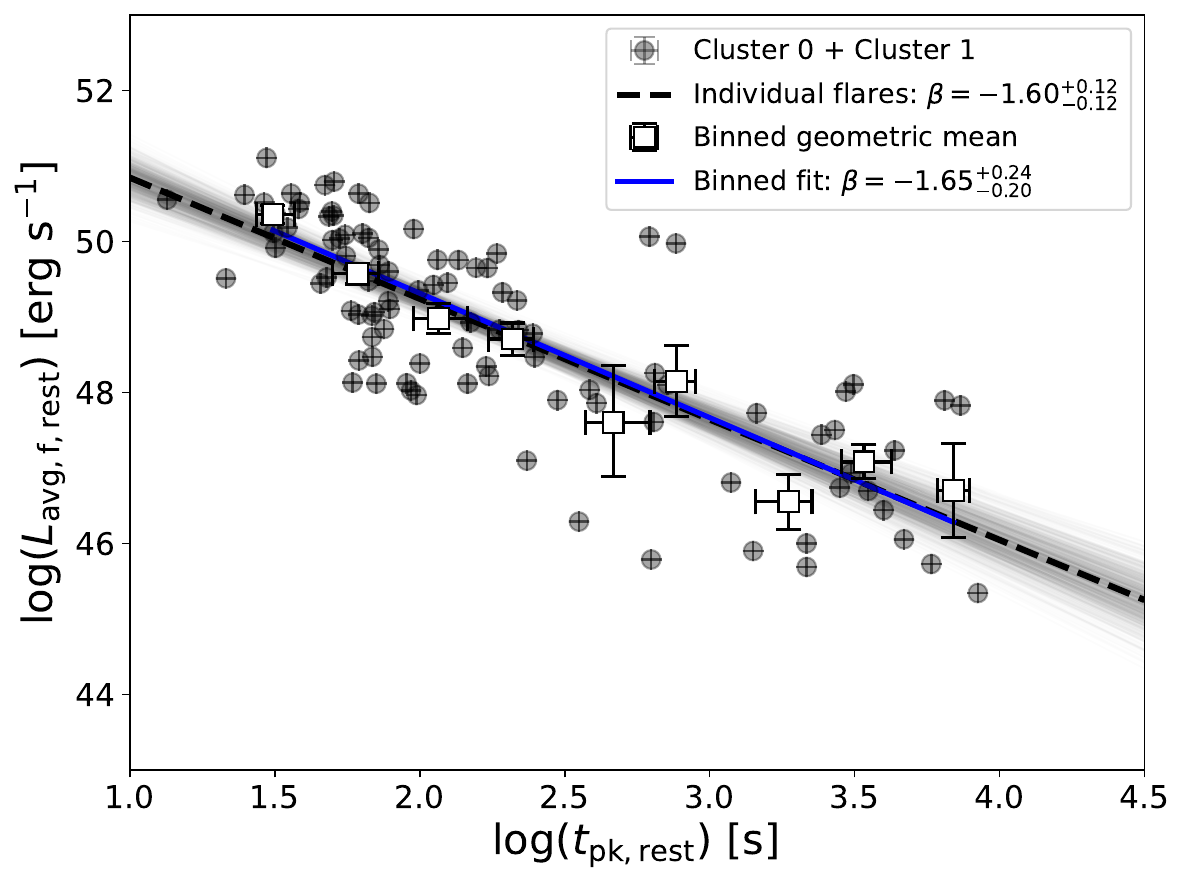}
\caption{Left: Relation between the average flare luminosity,
$L_{\rm avg,f,rest}=E_{\rm iso,f,rest}/w_{\rm rest}$, and the rest-frame flare peak time,
$t_{\rm pk,rest}$, for the early--narrow (purple) and late--wide (green) flare populations. The dashed lines show the median Bayesian linear regression obtained with {\tt LinMix}, while the faint lines represent random posterior realizations of the fitted relation. The quoted slopes correspond to the median posterior values with their 68\% credible intervals. A statistically significant anti-correlation is found for the early--narrow flares, whereas no significant trend is detected for the late--wide population. Because the average flare luminosity is not identical to the instantaneous jet luminosity, this relation provides only an indirect comparison with the luminosity evolution predicted by accretion-powered central-engine models. Right: Average flare luminosity as a function of rest-frame peak time for the combined flare sample (Cluster 0 + Cluster 1; grey circles), together with the geometric mean luminosities in logarithmic bins of peak time (open squares), following \citet{Margutti+2011}. The horizontal error bars on the binned measurements represent the 16th–84th percentile range of $\log(t_{\rm pk,rest})$ within each bin, while the vertical error bars indicate the bootstrap 68\% confidence intervals on the mean $\log(L_{\rm avg,f,rest})$. 
The black dashed line represents the Bayesian regression fitted to the individual flare measurements, with $\beta=-1.60\pm 0.12$, while the blue solid line shows the separate fit to the binned geometric mean luminosities, yielding $\beta=-1.65^{+0.24}_{-0.20}$. The two slopes are consistent within their uncertainties.}
\label{fig:scatter_Lavg_tpk}
\end{figure}

To provide an approximate comparison with the peak-luminosity prediction of episodic hyper-accretion models, we define the average flare luminosity as
\begin{equation}
L_{\rm avg,f,rest} = \frac{E_{\rm iso,f,rest}}{w_{\rm rest}},
\end{equation}
where $E_{\rm iso,f,rest}$ is the isotropic-equivalent energy released during the flare and $w_{\rm rest}$ is the flare width in the GRB rest frame. The corresponding luminosity–peak-time relations for the two flare populations are presented in the left panel of \ref{fig:scatter_Lavg_tpk} (left). We find a significant anti-correlation between the average flare luminosity and the flare peak time for the early--narrow flare population, following
$L_{\rm avg,f,rest}\propto
t_{\rm pk,rest}^{-1.74^{+0.24}_{-0.24}}$,
whereas no statistically significant correlation  
($L_{\rm avg,f,rest}\propto
t_{\rm pk,rest}^{-0.19^{+1.01}_{-0.93}}$)
is detected for the late--wide flares. Nevertheless, when the two populations are combined, the overall anti-correlation is preserved, yielding
$L_{\rm avg,f,rest}\propto t_{\rm pk,rest}^{-1.60^{+0.12}_{-0.12}}$.
This indicates that the global luminosity evolution is primarily driven by the early--narrow flares, while the late--wide flares introduce additional scatter rather than erasing the overall trend.

Although the observed slope is shallower than the
$L_{\rm pk}\propto t_{\rm pk}^{-3}$ prediction of episodic hyper-accretion models \citep{Kumar+2008a}, our analysis is based on the average flare luminosity rather than the peak luminosity. Consequently, a direct quantitative comparison is not yet possible. The dominance of the early--narrow flare population in establishing the observed luminosity evolution suggests that these events preserve the expected temporal evolution of the central engine, whereas the late--wide flares may originate from a different accretion regime or involve additional physical processes that weaken this relation.

%%%%%%%%%%%%%%%%%%%%%%%%%%%%%%%%%%%%%%%%%%%%%%%%%%%%%%%%%%
%%%%%%%% Figure: average flare luminosity and tpk,rest
%%%%%%%%%%%%%%%%%%%%%%%%%%%%%%%%%%%%%%%%%%%%%%%%%%%%%%%%%%
%number of words 
%%%%%%%%%%%%%%%%%%%%%%%%%%%%%%%%%%%%%%%%%%%%%%%%%%%%%%%%%%
%%%%%%%% 
%%%%%%%%%%%%%%%%%%%%%%%%%%%%%%%%%%%%%%%%%%%%%%%%%%%%%%%%%%

\subsubsection{Correlations between flare parameters: Flare time and flux variabilities} 
\label{sec:correlations3}

\begin{figure}%[ht!]
\centering
\includegraphics[width=0.9\linewidth]{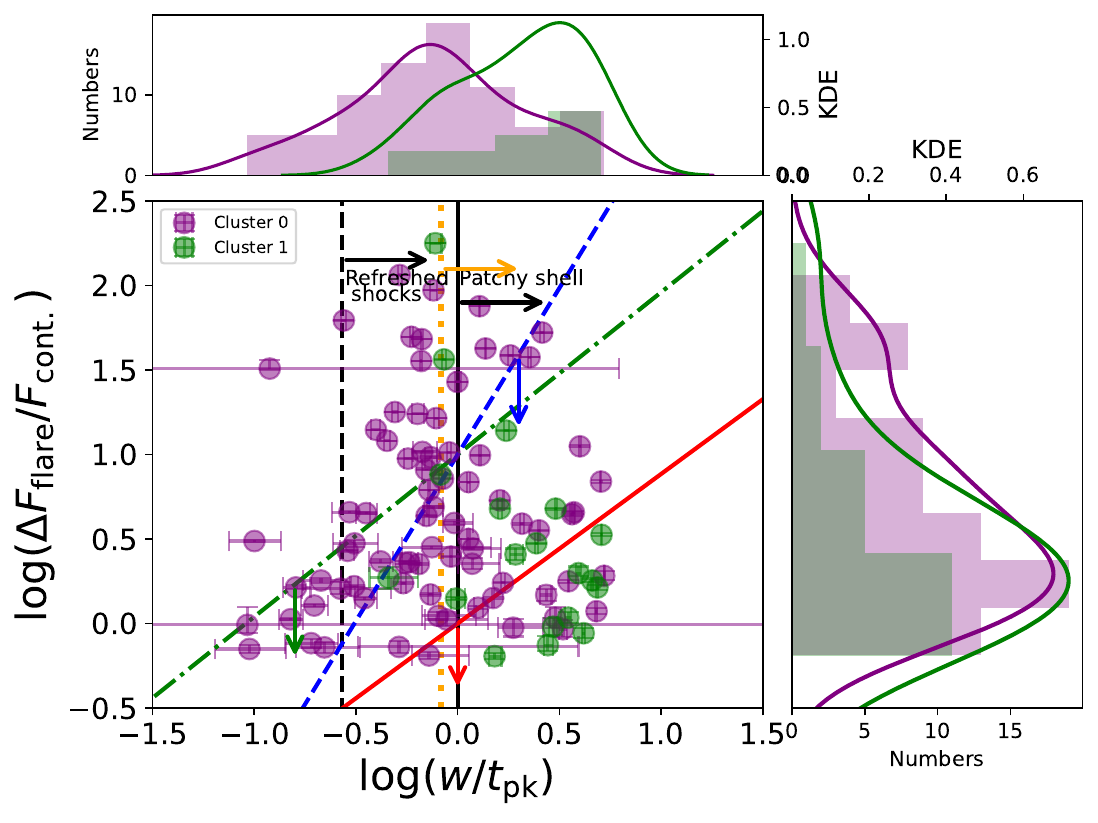}
\caption{The relation between the ratio of flare width to flare peak time, $\log(w/t_{\rm pk})$, and the flare flux variability relative to the underlying continuum, $\log(\Delta_{F,f}/F_{\rm cont.})$. Purple points represent the 78 early--narrow flares in Cluster~0, while green points represent the 19 late--wide flares in Cluster~1. Error bars represent the uncertainties propagated into logarithmic space. The distributions of these parameters, with the same color coding, are shown in the top and left panels, respectively. In each panel, the kernel density estimate (KDE) is shown by the purple and green solid lines for the corresponding samples.
The three limits, shown with solid red, dashed blue (representing bumps due to density fluctuations for the on-axis and off-axis cases, respectively), and dot-dashed green lines (representing simultaneous fluctuations in many regions for the off-axis case), indicate the kinematically allowed regions for afterglow variabilities. These regions have been computed using Equations~(7) and (A2) of \citet{IKZ05}, with the downward arrows indicating the corresponding allowed regions below each limit. Additionally, the solid and dashed vertical black lines represent the limits $\Delta t/t>1$ for bumps caused by patchy shells and $\Delta t/t>0.25$ for bumps caused by refreshed shocks, respectively, where $\Delta t$ is the variability timescale and $t$ is the observed time since the burst. The dotted orange vertical line indicates the lower limit $w/t_{\rm pk}\simeq0.83$ derived by \citet{LP07} for the shortest variability timescale expected from interactions with the circumburst medium in a wind environment. The rightward arrows associated with the vertical limits indicate the corresponding allowed regions toward larger $w/t_{\rm pk}$.
The Spearman rank correlation coefficients are $r=0.22$ ($p=0.28$) and $r=-0.37$ ($p=0.12$) for the early--narrow and late--wide flare populations (Clusters~0 and 1), respectively, indicating no statistically significant correlation in either subsample.}
\label{fig:hist_time_flux_varabilities}
\end{figure}

Figure~\ref{fig:hist_time_flux_varabilities} shows the relation between the ratio of flare width to flare peak time, $w/t_{\rm pk}$, and the flare flux variability relative to the underlying continuum, $\Delta_{F,f}/F_{\rm cont.}$. The distributions of these quantities are shown in the top and left panels, respectively. The early--narrow flares (Cluster~0) have an average value of $\langle \log(w/t_{\rm pk})\rangle=-0.12\pm0.05$ ($\sigma = 0.43$), corresponding to a characteristic variability timescale of $w/t_{\rm pk}\approx0.76^{+0.09}_{-0.08}$ ($1\sigma$ range: $0.29$--$2.04$), whereas late--wide flares (Cluster~1) have $\langle \log(w/t{\rm pk})\rangle=0.31\pm0.07$ ($\sigma = 0.31$), corresponding to $w/t_{\rm pk}\approx2.04^{+0.37}_{-0.31}$ ($1\sigma$ range: $0.99$--$4.20$). The difference is considerably smaller for the flux variability, with $\langle\log(\Delta{F,f}/F_{\rm cont.})\rangle=0.66\pm0.07$ ($\sigma = 0.59$) and $0.50\pm0.14$ ($\sigma = 0.62$) for the early--narrow and late--wide flare populations, respectively.

We also performed K--S tests for each parameter and found $D=0.51$ and $p=4\times10^{-4}$ for $w/t_{\rm pk}$, and $D=0.19$ and $p=0.59$ for $\Delta_{F,f}/F_{\rm cont.}$, indicating that the two flare populations differ significantly in their variability timescales, whereas no significant difference is found in their flux variability.

%%%%%%%%%%%%%%%%%%%%%%%%%%%%%%%%%%%%%%%%%%%%%%%%%%%%%%%%%
%%%%%%% Figure: w / tpk and DF/F histograms and correlation
%%%%%%%%%%%%%%%%%%%%%%%%%%%%%%%%%%%%%%%%%%%%%%%%%%%%%%%%%
%%%%%%%%%%%%%%%%%%%%%%%%%%%%%%%%%%%%%%%%%%%%%%%%%%%%%%%%%
%%%%%%% 
%%%%%%%%%%%%%%%%%%%%%%%%%%%%%%%%%%%%%%%%%%%%%%%%%%%%%%%%%

Finally, we find no statistically significant correlation between the flare time variability, $\log(w/t_{\rm pk})$, and the flux variability, $\log(\Delta F_{\rm flare}/F_{\rm cont.})$, in either flare population. For GRBs with early--narrow flares (Cluster 0), the Spearman rank correlation coefficient is $r=0.22$ ($p=0.28$), indicating at most a weak positive trend. For GRBs with late--wide flares (Cluster 1), we obtain $r=-0.37$ ($p=0.12$), corresponding to a weak negative trend. However, the relatively large chance probabilities indicate that neither trend is statistically significant.

\subsubsection{Normalized temporal properties and flare asymmetry}
\label{sec:normalized_temporal_properties}
\begin{figure}%[H]%[ht!]
\centering
\includegraphics[width=0.45\linewidth]{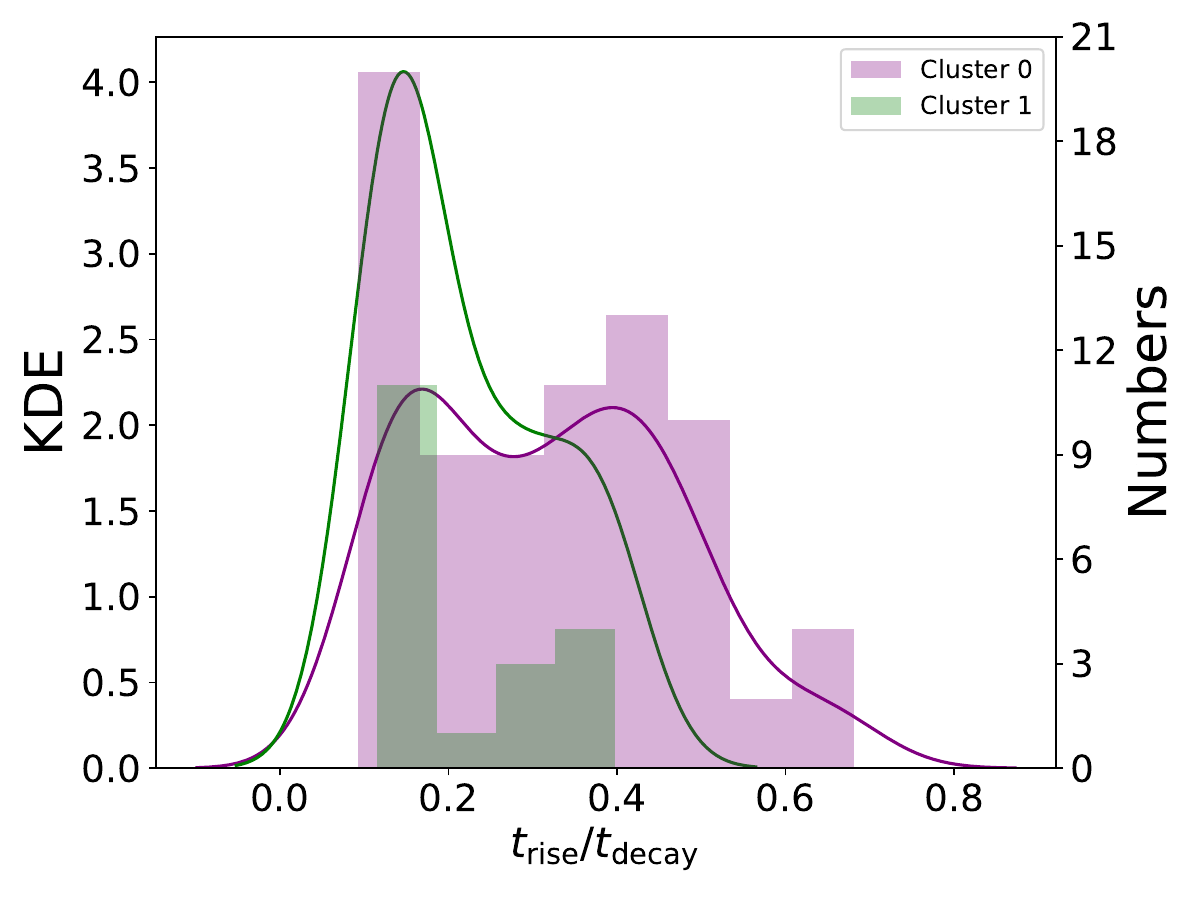}
\includegraphics[width=0.45\linewidth]{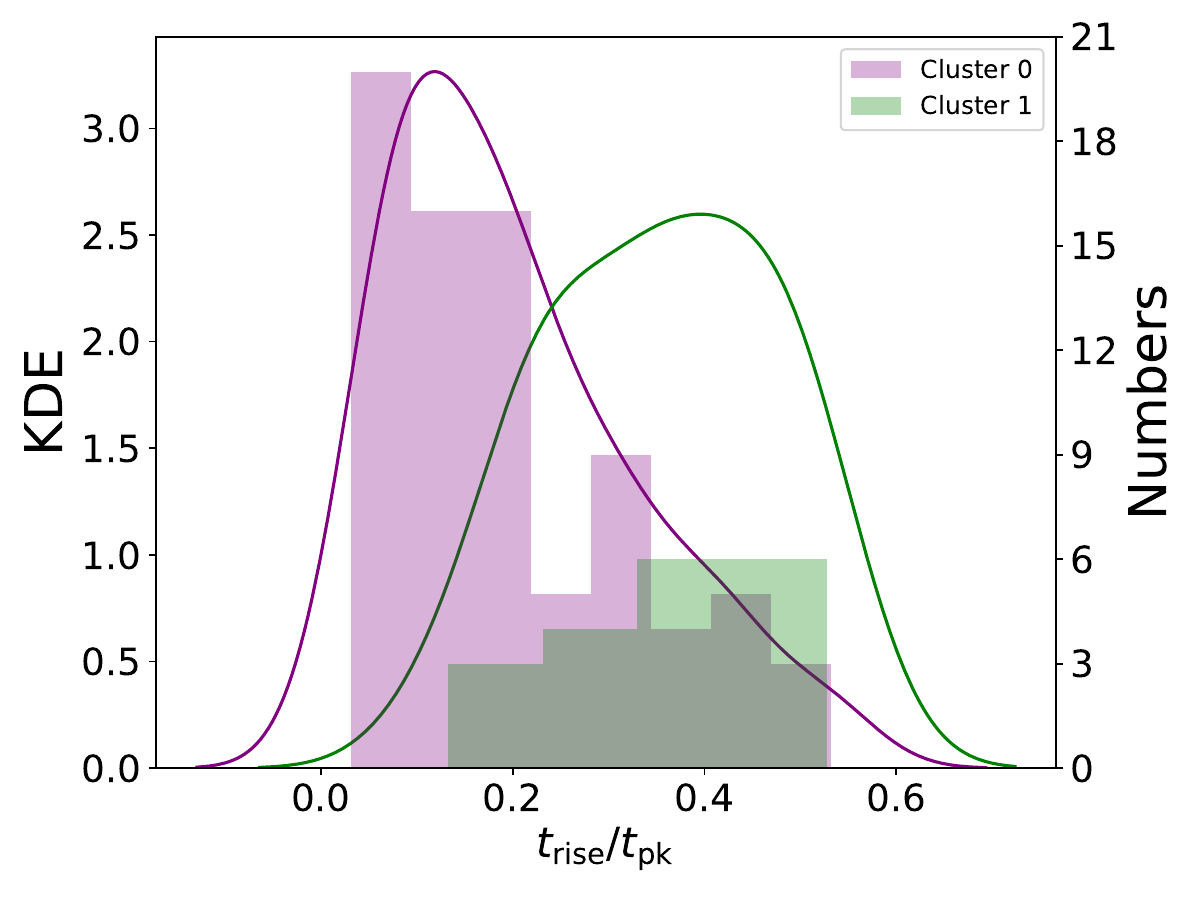}
\includegraphics[width=0.45\linewidth]{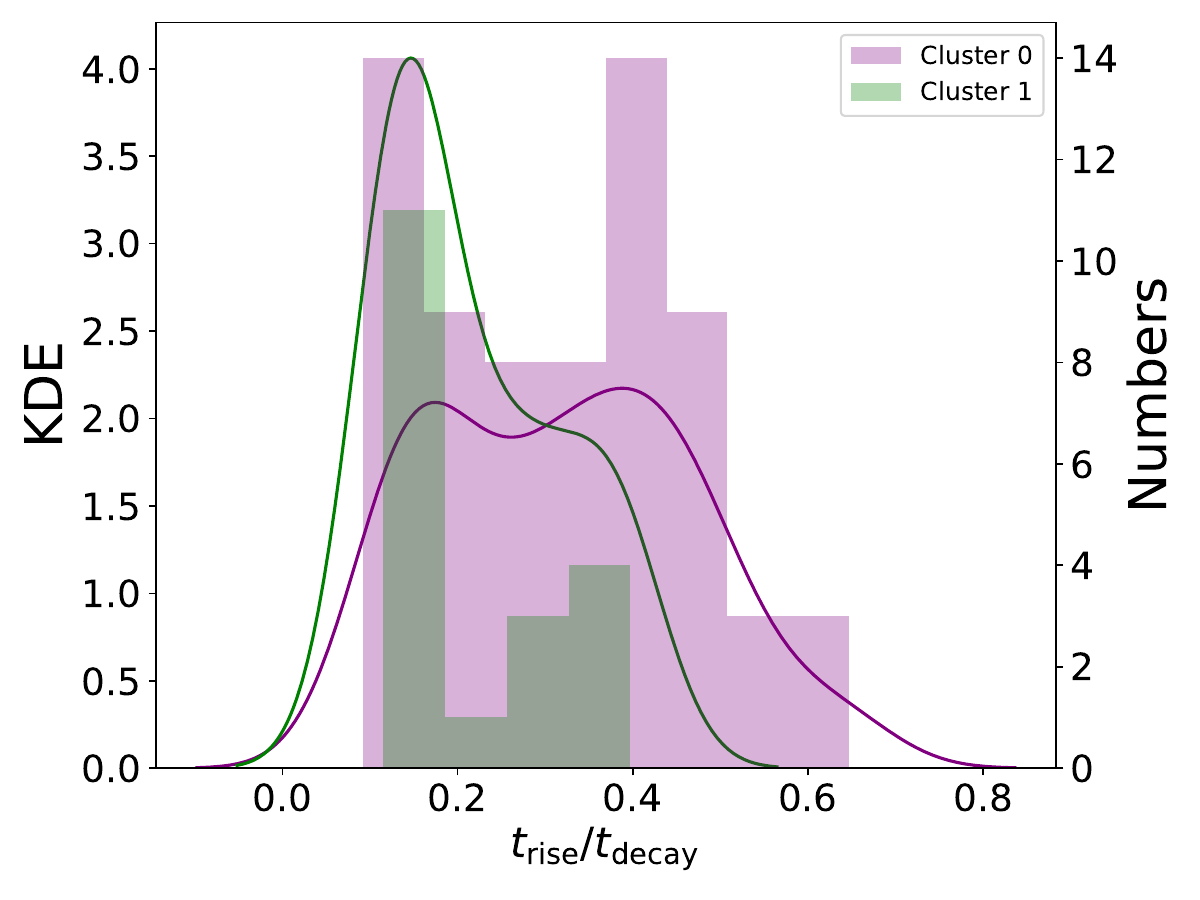}
\includegraphics[width=0.45\linewidth]{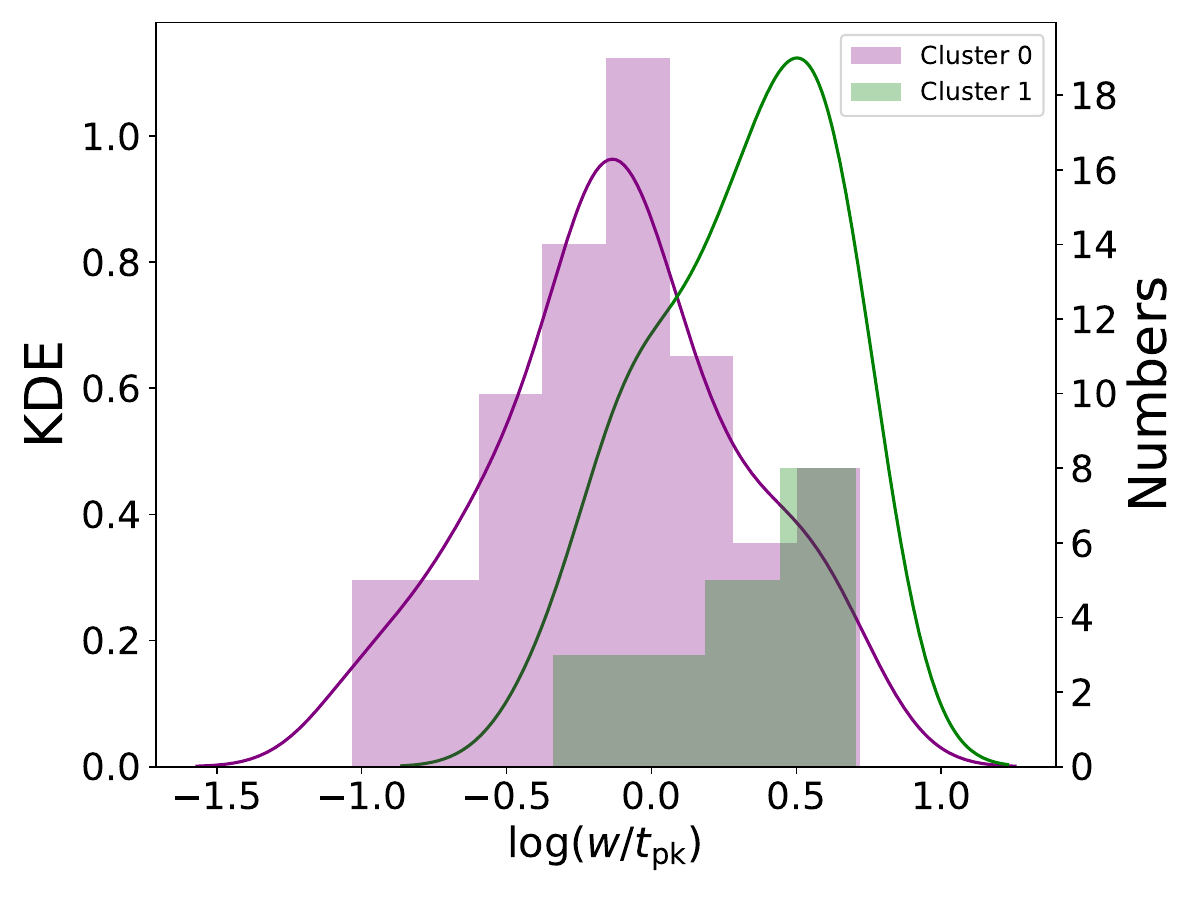}
\caption{
Top left: Distributions of the flare asymmetry, $t_{\rm rise}/t_{\rm decay}$. The early--narrow flare population
exhibits a possible bimodal structure, whereas the late--wide flare population is characterized by a dominant component with a tail extending toward larger values.
Bottom left: The same distributions after excluding potentially
prompt-contaminated flares ($T_{90}>t_{\rm pk}$). The overall behavior remains unchanged after excluding these flares, see Appendix~\ref{promt_effect}.
Top right: Distributions of the normalized rise time,
$t_{\rm rise}/t_{\rm pk}$. The early--narrow flares are preferentially concentrated at smaller values, whereas the late--wide flares are shifted toward larger values.
Bottom right: Distributions of the normalized flare width,
$w/t_{\rm pk}$. The early--narrow flares are concentrated at smaller values and exhibit a tail toward low $w/t_{\rm pk}$, whereas the late--wide distribution extends toward substantially larger values. In each panel, the right-hand ordinate shows the number of flares in each bin, while the left-hand ordinate shows the corresponding kernel density estimate (KDE).}
\label{fig:flare-rise_decay_ratio}
\end{figure}

The distributions of $t_{\rm rise}/t_{\rm decay}$ shown in
Figure~\ref{fig:flare-rise_decay_ratio} (top-left) indicate that both flare populations are dominated by asymmetric profiles with
$t_{\rm rise}<t_{\rm decay}$. The distribution of the early--narrow flares shows evidence for a possible bimodal structure within this population, distinct from the separation between the early--narrow and late--wide flare populations.
A two-component univariate GMM identifies components centered at
$t_{\rm rise}/t_{\rm decay}=0.14~(\sigma =0.02)$ and $0.38~(\sigma =0.12)$. In contrast, the late--wide flare distribution is visually dominated by a single broad peak with a tail extending toward higher values of $t_{\rm rise}/t_{\rm decay}$. \footnote{Interestingly, the more asymmetric component of the early--narrow flare population, centered at $t_{\rm rise}/t_{\rm decay}=0.14~(\sigma =0.02)$, lies close to the dominant asymmetry observed in the late--wide flare distribution.} The bimodal structure and the inferred GMM components remain essentially unchanged after excluding prompt-contaminated flares, as presented in Appendix~\ref{promt_effect}.

The early--narrow flares have an average value of
$\langle t_{\rm rise}/t_{\rm decay}\rangle=0.32\pm0.02$ ($\sigma=0.15$) and span a
relatively broad range up to $t_{\rm rise}/t_{\rm decay}\sim0.65$.
In contrast, the late--wide flares have a smaller average value of
$\langle t_{\rm rise}/t_{\rm decay}\rangle=0.22\pm0.02$ ( $\sigma=0.10$), with most values lying below $\sim0.4$. Thus, the late--wide flares have, on average, smaller $t_{\rm rise}/t_{\rm decay}$ ratios than the early--narrow flares.

In addition to the above investigation, we examined the distribution of the $t_{\rm rise}/t_{\rm pk}$ ratio for the two flare populations (Figure~\ref{fig:flare-rise_decay_ratio}, top-right). The early--narrow flares are preferentially concentrated at smaller values, with $\langle t_{\rm rise}/t_{\rm pk}\rangle=0.20\pm0.01$ ($\sigma=0.13$) , whereas the late--wide flares are shifted toward larger values, with $\langle t_{\rm rise}/t_{\rm pk}\rangle=0.36\pm0.03$ ($\sigma=0.12$) . The two distributions differ significantly, as confirmed by the Kolmogorov--Smirnov test ($D=0.56$, $p=4\times10^{-5}$).

The $w/t_{\rm pk}$ distributions
(Figure~\ref{fig:flare-rise_decay_ratio}, bottom-right) have
characteristic values of $\langle w/t_{\rm pk}\rangle \approx 0.76^{+0.09}_{-0.08}$ ($1\sigma$ range: $0.29$--$2.04$) and $\langle w/t_{\rm pk}\rangle \approx 2.04^{+0.37}_{-0.31}$ ($1\sigma$ range: $0.99$--$4.20$) for the early--narrow and late--wide flare populations, respectively. 
The K--S test yields $D=0.51$ and $p=4\times10^{-4}$, confirming
that the two distributions are statistically distinct. The
early--narrow distribution also exhibits a tail toward smaller
$w/t_{\rm pk}$ values, whereas the late--wide flares extend toward substantially larger values.

%%%%%%%%%%%%%%%%%%%%%%%%%%%%%%%%%%%%%%%%%%%%%%%%%%%%%%%%%%
%%%%%%%% Figure: ratios of trise/tdecay, trise/tpk
%%%%%%%%%%%%%%%%%%%%%%%%%%%%%%%%%%%%%%%%%%%%%%%%%%%%%%%%%%
%number of words
%%%%%%%%%%%%%%%%%%%%%%%%%%%%%%%%%%%%%%%%%%%%%%%%%%%%%%%%%%
%%%%%%%%
%%%%%%%%%%%%%%%%%%%%%%%%%%%%%%%%%%%%%%%%%%%%%%%%%%%%%%%%%%

Finally, we examined the temporal evolution of the flare asymmetry. For the early--narrow population, both the rise and decay times show significant positive correlations with flare peak time ($r_s=0.80$, $p=1.6\times10^{-18}$ and $r_s=0.67$, $p=2.1\times10^{-11}$, respectively). 
For the late--wide population, the rise time is also strongly correlated with peak time ($r_s=0.77$, $p=1.0\times10^{-4}$), whereas the correlation between decay time and peak time is weaker and not statistically significant ($r_s=0.41$, $p=0.082$).
In contrast, we find no statistically significant correlation between the flare asymmetry ($t_{\rm rise}/t_{\rm decay}$) and $t_{\rm pk}$ for either population ($r_s=-0.04$, $p=0.75$ for the early--narrow flares and $r_s=0.18$, $p=0.47$ for the late--wide flares).

\subsection{Waiting times between successive flare episodes}

\label{Waiting-time_distributions}

\begin{figure}%[ht!]
\centering
\includegraphics[width=0.8\linewidth]{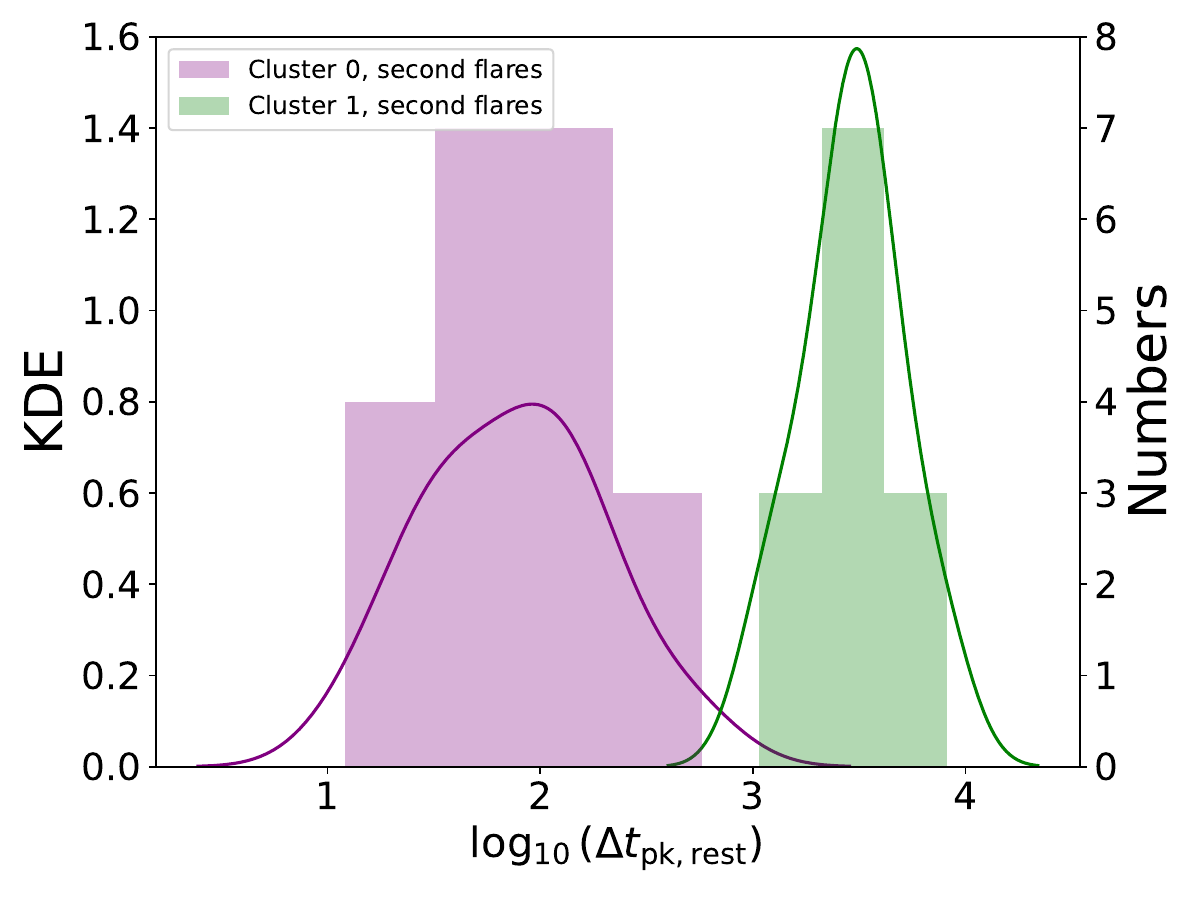}
\caption{
Distribution of the rest-frame waiting time between the first and second X-ray flares, defined as
$\Delta t_{\rm pk,rest}=t_{\rm pk,2,rest}-t_{\rm pk,1,rest}$,
for GRBs in which the second flare belongs to the early--narrow (Cluster~0; purple) or late--wide (Cluster~1; green) population. 
The early--narrow population has a mean waiting time of
$\log_{10}(\Delta t_{\rm pk,rest}/{\rm s})=1.88\pm0.09$, $\sigma = 0.42$
($\Delta t_{\rm pk,rest}\approx76~{\rm s}$; $N=21$),
whereas the late--wide population has a mean waiting time of
$\log_{10}(\Delta t_{\rm pk,rest}/{\rm s})=3.47\pm0.06$, $\sigma = 0.24$
($\Delta t_{\rm pk,rest}\approx2950~{\rm s}$; $N=13$).
The two populations exhibit distinct characteristic waiting-time scales, with only limited overlap between their distributions.}
\label{fig:hist_waiting-time_second_flares}
\end{figure}

Having established that the early--narrow and late--wide flare
populations exhibit different temporal properties, we next investigate whether this distinction also extends to the time intervals between successive flare episodes. We therefore examine the rest-frame waiting time between successive flares in GRBs exhibiting two flare episodes, as presented in Figure~\ref{fig:hist_waiting-time_second_flares}.
We define the waiting time as
\[
\Delta t_{\rm pk,rest}=t_{{\rm pk},2,{\rm rest}}-t_{{\rm pk},1,{\rm rest}},
\]
where $t_{{\rm pk},1,{\rm rest}}$ and $t_{{\rm pk},2,{\rm rest}}$ are the rest-frame peak times of the first and second flares, respectively.\footnote{Two GRBs (170113A and 141221A) were excluded from this analysis because the fitted peak time of the second flare component was slightly earlier than that of the first flare, resulting in negative values of $\Delta t_{\rm pk,rest}$. For GRB~170113A, $t_{\rm pk,1,rest}=31.79$~s and $t_{\rm pk,2,rest}=31.30$~s ($\Delta t_{\rm pk,rest}=-0.48$~s), while for GRB~141221A, $t_{\rm pk,1,rest}=146.19$~s and $t_{\rm pk,2,rest}=140.69$~s ($\Delta t_{\rm pk,rest}=-5.50$~s). In these cases, the first flare component begins earlier but has a longer rise time than the second component, allowing the later-starting component to reach its peak slightly earlier and thereby reversing the chronological order of the fitted peak times. Because our waiting-time analysis requires chronologically ordered flare peaks, these two events were excluded.} The classification is based on the properties of the second flare. Therefore, only GRBs exhibiting two flare episodes are included in this analysis, resulting in 21 early--narrow (Cluster~0) and 13 late--wide (Cluster~1) second flares.

We find that the rest-frame waiting-time distributions differ significantly between the two flare populations. The early--narrow second flares have a mean waiting time of $\log_{10}(\Delta t_{\rm pk,rest}/{\rm s}) = 1.88 \pm 0.09$ ($\sigma = 0.42$), corresponding to a characteristic waiting time of $\sim76$~s, whereas the late--wide second flares have a mean waiting time of $\log_{10}(\Delta t_{\rm pk,rest}/{\rm s}) = 3.47 \pm 0.06$ ($\sigma = 0.24$), corresponding to $\sim2950$~s. A Kolmogorov--Smirnov test yields $D=1.00$ and $p=2.16\times10^{-9}$, indicating that the two waiting-time distributions are highly unlikely to originate from the same parent population.

%%%%%%%%%%%%%%%%%%%%%%%%%%%%%%%%%%%%%%%%%%%%%%%%%%%%%%%%%%
%%%%%%%% Figure: Normalized waiting time between successive flares
%%%%%%%%%%%%%%%%%%%%%%%%%%%%%%%%%%%%%%%%%%%%%%%%%%%%%%%%%%
%number of words
%%%%%%%%%%%%%%%%%%%%%%%%%%%%%%%%%%%%%%%%%%%%%%%%%%%%%%%%%%
%%%%%%%%
%%%%%%%%%%%%%%%%%%%%%%%%%%%%%%%%%%%%%%%%%%%%%%%%%%%%%%%%%%

%%%%%%%%%%%%%%%%%%%%%%%%%%%%%%%%%%%%%%%%%%%%%%%%%%%%%%%%%%
%%%%%%%% 
%%%%%%%%%%%%%%%%%%%%%%%%%%%%%%%%%%%%%%%%%%%%%%%%%%%%%%%%%%

\section{Discussion}
\label{sec:discussion}

The statistical identification of two flare populations, early--narrow and late--wide, provides a new framework for interpreting the temporal evolution of GRB X-ray flares. Beyond their distinct locations in the peak-time--width plane, the two populations exhibit differences in their luminosity evolution, temporal variability, waiting times, and flare asymmetry. These properties provide new constraints on whether X-ray flares can be described as a single population evolving continuously with time and on the physical processes responsible for late-time GRB activity.

In the following discussion, we first discuss the temporal properties of the two flare populations, including their asymmetry, characteristic variability timescales, and compare their temporal behavior with that of prompt-emission pulses. We then examine the flare time and flux variability in the context of external-shock constraints. Finally, we explore the implications of these observational properties for accretion-powered central-engine models.

\subsection{Temporal evolution of the two flare populations}
\label{dis:temporal_properties}
%\label{dis:bimodal-distributions}

One of the main temporal properties distinguishing the two flare populations is their asymmetry. Both populations are dominated by profiles with $t_{\rm rise}<t_{\rm decay}$, although the late--wide flares have, on average, smaller $t_{\rm rise}/t_{\rm decay}$ ratios ($0.22\pm0.02$, $\sigma=0.10$) than the early--narrow flares ($0.32\pm0.02$, $\sigma=0.15$), where the quoted uncertainties represent the standard errors of the mean (SEM) and $\sigma$ denotes the standard deviation of each distribution. This difference indicates that the two populations differ not only in their absolute temporal scales but also in their temporal asymmetry. Previous studies generally reported a single broad distribution of the X-ray flare asymmetry ratio for flares before $t_{\rm pk}<1000s$ \citep{Chincarini+10}. In contrast, after separating the flares into the early--narrow and late--wide populations, we find evidence for an additional possible bimodal structure within the early--narrow population. The temporal asymmetry of the early--narrow flares may therefore not be described by a single underlying distribution.

The average asymmetry ratios of both populations are lower than the characteristic values previously reported for prompt-emission pulses, for which \citet{KRL03} found a median $t_{\rm rise}/t_{\rm decay} =0.47\pm0.01$ ($\sigma=0.09$), and for early X-ray flares, for which \citet{Chincarini+10} reported $\langle t_{\rm rise}/t_{\rm decay}\rangle=0.49$ ($\sigma = 0.26$). More recent studies of prompt emission, however, have also found
increasingly asymmetric emission episodes. \citet{Hakkila+18} showed that bright, complex GRB emission episodes can be strongly asymmetric, while \citet{Gowri+25} found that later pulses in multi-pulse GRBs observed by \textit{Fermi}-GBM tend to exhibit stronger asymmetries. Because these studies adopt different definitions of pulse asymmetry, their reported asymmetry parameters cannot be directly compared quantitatively with $t_{\rm rise}/t_{\rm decay}$ used here. Interestingly, we observe a qualitatively similar tendency, with the later, late--wide flares exhibiting, on average, stronger asymmetries than the earlier, early--narrow flares. This behavior is broadly consistent with the increasing asymmetry reported for prompt-emission episodes occurring at later times.

Although the flare rise and decay times generally increase with flare peak time, their ratio shows no significant evolution with $t_{\rm pk}$ within either population. This indicates that the characteristic flare asymmetry is approximately preserved as the absolute temporal scales increase within each population. Such behavior differs from the temporal evolution reported for prompt-emission pulses, for which early pulses are relatively symmetric, whereas later pulses become progressively more
asymmetric \citep{Gowri+25}. Although recent studies have suggested a possible continuity between late prompt-emission pulses and X-ray flares \citep{Hakkila+24,Gowri+25}, our results indicate that the flare asymmetry does not exhibit a comparable systematic evolution with $t_{\rm pk}$.

Previous studies of X-ray flares have similarly shown that the rise and decay times evolve in a coupled manner, while their asymmetry exhibits little evolution with flare time \citep[e.g.,][]{Peng+15}. Our results extend this picture by showing that the characteristic flare asymmetry remains approximately preserved with peak time separately within the early--narrow and late--wide populations, despite the different average asymmetries and markedly different temporal scales of the two populations. In addition, \citet{Sonbas+13} found that the minimum variability timescale (MVTS) is tightly correlated with the flare rise time, while \citet{SIM23} reported a positive correlation between the MVTS and flare duration. Together with the observed correlations between flare peak time and the characteristic rise and decay times in our sample, these results suggest that several temporal properties of X-ray flares, including the rise time, decay time, flare width, peak time, and minimum variability timescale, are interconnected rather than evolving independently.

The normalized rise time, $t_{\rm rise}/t_{\rm pk}$, provides an
additional distinction between the two flare populations. The
early--narrow flares have systematically smaller $t_{\rm rise}/t_{\rm pk}$ ratios ($0.20\pm0.01$, $\sigma=0.13$) than the late--wide flares ($0.36\pm0.03$, $\sigma=0.12$), and the two distributions are statistically distinct. This demonstrates that
the difference between the two populations is not limited to their absolute peak times and widths, but also extends to their normalized temporal structure. In particular, the late--wide flares cannot be interpreted simply as early--narrow flares shifted to later times while preserving the same relative rise timescale.

The flare populations also differ significantly in their normalized widths. The characteristic $w/t_{\rm pk}$ ratio increases from $0.76^{+0.09}_{-0.08}$ ($1\sigma$ range: $0.29$--$2.04$) for the early--narrow flares to $2.04^{+0.37}_{-0.31}$ ($1\sigma$ range: $0.99$--$4.20$) for the late--wide flares. Together with the different $t_{\rm rise}/t_{\rm pk}$ distributions, this demonstrates that the distinction between the two populations is not simply a consequence of the late--wide flares occurring at later times and having larger absolute widths. Their characteristic durations also scale differently relative to their peak times.

The correlations between flare peak time and the characteristic flare timescales are generally stronger for the early--narrow population, whereas some of these relations become weaker for the late--wide flares (Section~\ref{sec:normalized_temporal_properties}). In particular, while the rise time remains strongly correlated with peak time in the late--wide population, the corresponding decay-time relation is weaker and not statistically significant. Together with the differences in their normalized temporal properties, this indicates that the late--wide flares are not simply a temporally stretched continuation of the early--narrow population.

Taken together, the distributions of $t_{\rm rise}/t_{\rm decay}$, $t_{\rm rise}/t_{\rm pk}$, and $w/t_{\rm pk}$, together with the different temporal scaling relations exhibited by the two flare populations, indicate that the early--narrow and late--wide flares are unlikely to represent a single self-similar population evolving only in time. Instead, they appear to exhibit distinct modes of temporal evolution.

\subsection{Are X-ray flares distinct from prompt-emission pulses?}
\label{correlations4}

Previous studies have also investigated the possible continuity
between prompt-emission pulses and X-ray flares through their waiting
times. \citet{Guidorzi+15} showed that the overall waiting-time distribution extending from prompt $\gamma$-ray pulses to X-ray flares, up to approximately $10^{3}$ s for GRBs with known redshifts, can be described by a common time-dependent Poisson process. Their
analysis considered the waiting times of successive emission episodes
collectively, without distinguishing different X-ray flare
populations. Motivated by this result, we examined whether the two
flare populations identified in our analysis exhibit similar
waiting-time behavior.

In contrast to considering all X-ray flare waiting times together, we
find that conditioning the waiting time on the temporal properties of
the second flare reveals a pronounced difference between the two
populations. As shown in
Figure~\ref{fig:hist_waiting-time_second_flares}, late--wide second
flares are preceded by substantially longer rest-frame waiting times
than early--narrow second flares, with characteristic waiting times of
$\sim2950$~s and $\sim76$~s, respectively. Thus, our result does not contradict the global stochastic behavior reported by \citet{Guidorzi+15}, but reveals additional structure within the X-ray flare population that is lost when waiting times are considered collectively, particularly at longer timescales beyond those covered by their analysis.
Although the waiting times are not independent of the peak times used in the classification, their distinct distributions suggest that the temporal continuity between prompt-emission pulses and X-ray flares may not characterize both flare populations in the same way.

\begin{figure}[ht!]
\centering
\includegraphics[width=0.8\linewidth]{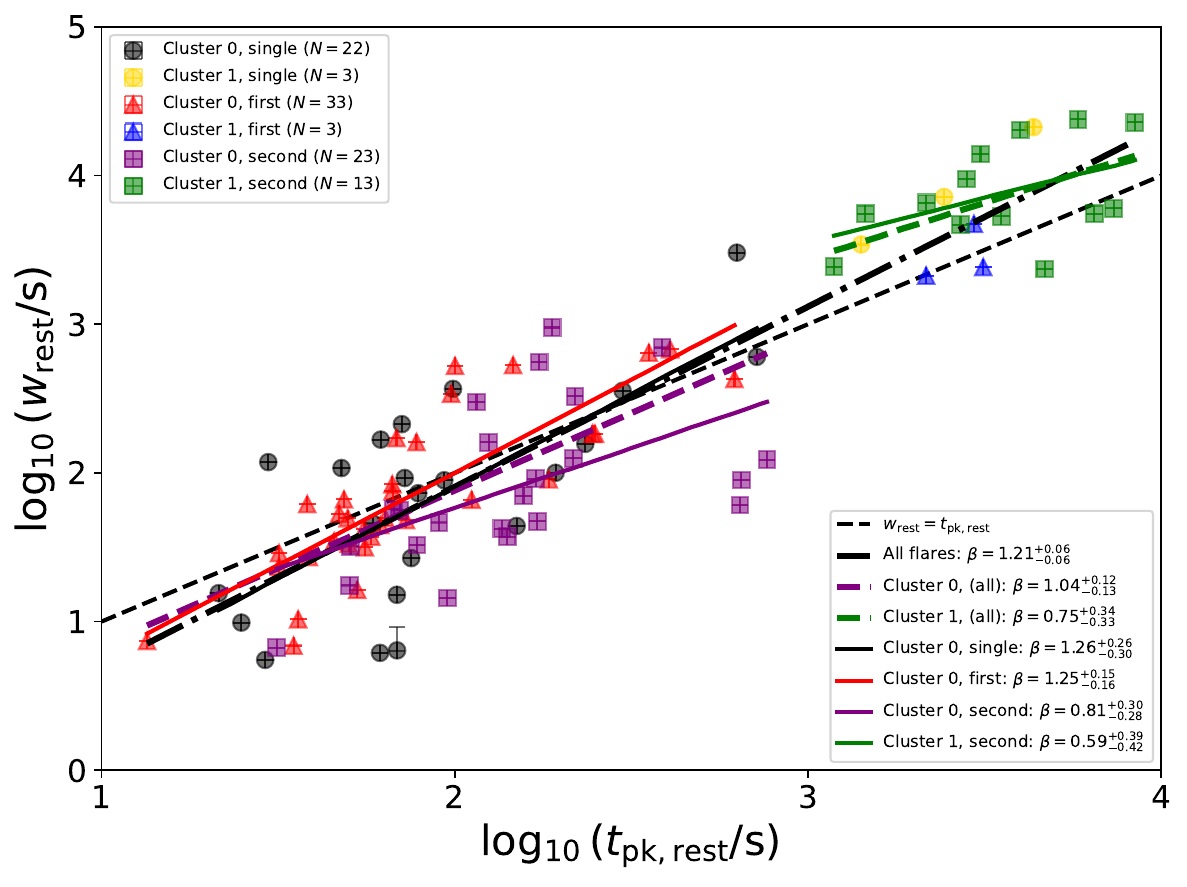}
\caption{ The relation between rest-frame flare peak time ($t_{\rm pk,rest}$) and width ($w_{\rm rest}$) for the 97 X-ray flares. The Gaussian Mixture Model identifies two populations: 78 early–narrow flares (Cluster 0) and 19 late–wide flares (Cluster 1). Different symbols distinguish single, first, and second flare episodes, while colors indicate the two populations. The solid lines represent the Bayesian linear regression fits to the complete sample and to each population separately, following the method of \citet{Kelly2007}. The dashed line indicates $w_{\rm rest}$ = $t_{\rm pk,rest}$.
Single and first flares predominantly belong to the early–narrow population (22 of 25 and 33 of 36, respectively). Although most late–wide flares are second episodes (13 of 19; 68\%), the majority of second flares remain in the early–narrow population (23 of 36; 64\%). The strong width–peak-time correlation observed for the combined sample is retained within the early–narrow population, whereas the corresponding correlation for the late–wide population is not statistically significant.}
\label{fig:scatter_tpk_width_second_flare}
\end{figure}

The possible connection between X-ray flares and prompt-emission
pulses can also be examined through their temporal scaling relations.
In particular, the width--peak-time correlation has previously been
considered a characteristic property of X-ray flares, in contrast to
prompt-emission pulses \citep{Chincarini+10}. We therefore examine the relation between flare width and peak time for the two populations (Cluster 0 and Cluster 1), further distinguishing single, first, and second flare episodes, as shown in Figure~\ref{fig:scatter_tpk_width_second_flare}. 
%%%%%%%%%%%%%%%%%%%%%%%%%%%%%%%%%%%%%%%%%%%%%%%%%%%%%%%%%%
%%%%%%%% Figure: correlation between tpk vs width (colored)
%%%%%%%%%%%%%%%%%%%%%%%%%%%%%%%%%%%%%%%%%%%%%%%%%%%%%%%%%%
%%%%%%%%%%%%%%%%%%%%%%%%%%%%%%%%%%%%%%%%%%%%%%%%%%%%%%%%%%
%%%%%%%% 
%%%%%%%%%%%%%%%%%%%%%%%%%%%%%%%%%%%%%%%%%%%%%%%%%%%%%%%%%%
When all flares are considered together, we recover the well-known strong positive correlation between the rest-frame flare width and peak time, with a Spearman rank coefficient of $r_s=0.84$ ($p=6.5\times10^{-27}$). A Bayesian linear regression following the method of \citet{Kelly2007}, applied to the relation $\log w_{\rm rest}=\alpha+\beta\log t_{\rm pk,rest}$, yields a slope of $\beta=1.21^{+0.06}_{-0.06}$. However, separating the sample into the early--narrow (Cluster~0) and late--wide (Cluster~1) populations reveals underlying differences that are not apparent in the combined sample. The early--narrow flares retain a strong correlation ($r_s=0.70$, $p=9.4\times10^{-13}$) and exhibit a nearly linear scaling, with $\beta=1.04^{+0.12}_{-0.13}$. In contrast, the late--wide flares exhibit only a moderate correlation ($r_s=0.45$, $p=0.055$), which falls just short of statistical significance, and their regression slope is substantially less well constrained, $\beta=0.75^{+0.34}_{-0.33}$~\footnote{The dependence on flare order does not alter this overall picture. Among the early--narrow flares, the width--peak-time correlation remains
strong for both first flare episodes ($r_s=0.86$,
$p=9.8\times10^{-11}$; $N=33$) and second flare episodes
($r_s=0.69$, $p=2.7\times10^{-4}$; $N=23$). The corresponding
Kelly regression slopes are $\beta=1.25^{+0.15}_{-0.16}$ and
$\beta=0.81^{+0.30}_{-0.28}$, respectively, and are consistent
within their uncertainties. For the late--wide population, the
first-flare subsample contains only three events and is therefore
too small to meaningfully assess the correlation. The second
late--wide flares ($N=13$) show no statistically significant
correlation ($r_s=0.37$, $p=0.21$), with a poorly constrained
Kelly regression slope of $\beta=0.59^{+0.39}_{-0.42}$.}.

These results show that the strong width--peak-time correlation observed for the full sample does not reflect identical behavior within the two flare populations. In particular, the early--narrow population exhibits a strong internal correlation with a regression slope consistent with unity, implying that $w/t_{\rm pk}$ remains approximately constant across this population. In contrast, the late--wide population exhibits a weaker correlation and a substantially less constrained scaling. Moreover, because late--wide flares preferentially occupy the region of larger $t_{\rm pk}$ and larger $w$, the separation between the two populations contributes to the particularly strong correlation observed when all flares are analyzed together. Thus, the global width--peak-time relation . reflects both the intrinsic scaling within the early--narrow population and the separation of the two populations in the $w$--$t_{\rm pk}$ plane, suggesting that the two populations may trace different regimes of flare evolution rather than a single uniform scaling behavior.

Previous studies interpreted the width--peak-time correlation as evidence that X-ray flares differ from prompt-emission pulses, which were not thought to exhibit a similar scaling \citep{Ramirez-Ruiz+2000, Chincarini+07, Chincarini+10, Bernardini+11, Yi+16}. More recent analyses of prompt GRB pulses, however, have suggested increasing similarities between late prompt pulses and X-ray flares \citep{Hakkila+24, Gowri+25}. Our results indicate that the correlation itself is not universal among flares, but is instead primarily associated with the early--narrow flare population, providing an additional observational distinction between the two classes identified in this work. Regardless of their possible connection to the prompt-emission phase, the origin of the X-ray flare variability itself remains to be established. We therefore next examine whether the observed temporal and flux variability can be accommodated within external-shock scenarios.

\subsection{Can X-ray flares have an external-shock origin?}
\label{Flare_time-flux_variability}
One of the fundamental questions in flare studies is whether these events can arise from variability associated with the external shock in the standard fireball model \citep{MZ09}. To explore this, \citet{IKZ05} theoretically investigated the relation between temporal variability ($w/t_{\rm pk}$, or relative width) and flux variability ($\Delta_{F,f}/F_{\rm cont}$) in GRB afterglows. They derived kinematic constraints on the allowed regions in the $\Delta_{F,f}/F_{\rm cont}$--$w/t_{\rm pk}$ parameter space for variability produced by external mechanisms, including density fluctuations, refreshed shocks, and patchy shells. The different constraints arise from the distinct geometrical and dynamical assumptions of these scenarios: density fluctuations correspond to variations in the ambient medium, patchy shells to angular structure of the emitting surface, and refreshed shocks to slower ejecta catching up with and re-energizing the external shock \citep{IKZ05}. These theoretical boundaries are overplotted in Figure~\ref{fig:hist_time_flux_varabilities}.

The early--narrow and late--wide flare populations are not clearly separated in this parameter space (Figure~\ref{fig:hist_time_flux_varabilities}); therefore, we consider the flare sample as a whole for this comparison. According to the kinematic constraints derived by \citet{IKZ05}, most of the flares in our sample are consistent with the regions allowed by one or more external-shock variability mechanisms. In particular, a substantial fraction of the flares is consistent with the patchy-shell (where  $w/t_{\rm pk}>1$) and refreshed-shock constraints, while fewer events are compatible with variability produced by density fluctuations. These regions correspond to different physical assumptions and therefore should be interpreted as kinematically allowed regions rather than as unique identifications of the flare-production mechanism.

Similar comparisons with these kinematic constraints were performed in previous X-ray flare studies, yielding broadly similar conclusions \citep{Chincarini+07,Chincarini+10,Bernardini+11}. Thus, when the $w/t_{\rm pk}$--$\Delta_{F,f}/F_{\rm cont}$ diagnostic is considered alone, our results, like those of previous studies, do not by themselves exclude an external-shock origin for the flares.

A complementary constraint was derived by \citet{LP07}, who showed that the shortest variability timescale expected from interactions with the circumburst medium is set by adiabatic expansion in a wind environment. For a spectral index of $\beta\simeq1$, they obtained a lower limit of $w/t_{\rm pk}\simeq0.83$ (their Fig.~2), which is overplotted in Figure~\ref{fig:hist_time_flux_varabilities}~\footnote{This comparison is particularly relevant to our previous analysis, in which the plateau phase is interpreted within a wind-like circumburst environment \citep{DPR22, DB+26}. We measured characteristic values of $w/t_{\rm pk}\simeq1.02^{+0.12}_{-0.11}$ ($1\sigma$ range: $0.42$--$2.49$) and $0.76^{+0.18}_{-0.15}$ ($1\sigma$ range: $0.23$--$2.55$) for GRBs with and without a plateau, respectively. The value for the plateau population therefore lies close to the \citet{LP07} limit derived for a wind environment.}. Approximately half of the flares in our sample (49/97; 50.5\%) lie above this limit and are therefore consistent with the variability timescale expected in this external-shock scenario. This fraction differs substantially between the two populations: 33/78 (42.3\%) of the early--narrow flares and 16/19 (84.2\%) of the late--wide flares lie above the \citet{LP07} limit. Correspondingly, the characteristic value of the early--narrow population, $w/t_{\rm pk}\simeq0.76^{+0.09}_{-0.08}$ ($1\sigma$ range: $0.29$--$2.04$), lies close to the \citet{LP07} limit, whereas the late--wide population has a substantially larger characteristic value, $w/t_{\rm pk}\simeq2.04^{+0.37}_{-0.31}$ ($1\sigma$ range: $0.99$--$4.20$).

However, recent multiwavelength observations provide a different perspective on this issue.  \citet{Tiwari+26} investigated 66 X-ray flares with simultaneous Swift-XRT and Fermi-LAT coverage and found significant GeV emission during five flare 
intervals. They showed that the detected GeV emission is consistent with the external continuous forward-shock afterglow origin rather than the X-ray flare component. 

Similarly, consistency with the kinematic constraints and the variability-timescale limit of \citet{LP07} does not by itself establish an external-shock origin.
Our analysis provides additional observational constraints on the nature of the flare emission that are not incorporated into these diagnostics. In particular, our previous analysis showed that the flare properties are largely independent of the presence of an X-ray plateau and the properties of the underlying afterglow \citep{DB+26}. Furthermore, the two flare populations exhibit significantly different normalized temporal properties. For example, the early--narrow flares have systematically smaller $t_{\rm rise}/t_{\rm pk}$ and $w/t_{\rm pk}$ ratios than the late--wide flares, with the normalized rise-time distributions differing significantly ($\langle t_{\rm rise}/t_{\rm pk}\rangle=0.20\pm0.01$, $\sigma = 0.13$ and $0.36\pm0.03$, $\sigma = 0.12$ respectively). If both populations arise from the same external-shock variability mechanism, this mechanism would therefore also need to account for the systematic difference in their normalized temporal structure. Thus, although an external-shock origin cannot be excluded from the variability constraints alone, the broader set of observational properties favors an internal origin associated with prolonged central-engine activity.

\subsection{Implications for accretion-powered central engine models}
\label{discussion_accretion}
One of the leading interpretations for the origin of X-ray flares is that they are powered by intermittent accretion onto the newly formed compact object following the prompt GRB. In this framework, the jet luminosity approximately traces the accretion power,
\begin{equation}
L_{\rm jet} \simeq \eta \dot{M}_{\rm acc}c^{2},
\end{equation}
where $\eta$ is the efficiency of converting accretion power into jet luminosity. As the available material surrounding the central engine is gradually depleted, both the accretion rate and the flare luminosity are expected to decrease with time \citep{King+2005, Perna+2006, Proga&Zhang2006, Kumar+2008a}.
Several physical mechanisms have been proposed to produce the intermittent accretion responsible for X-ray flares, including gravitational fragmentation of the accretion disk \citep{Perna+2006}, fragmentation of fallback material feeding the disk \citep{Kumar+2008a}, and magnetic barriers that periodically halt and restart the accretion flow \citep{Proga&Zhang2006}.

The temporal evolution of the jet luminosity provides an important theoretical prediction of accretion-powered central-engine models. In the fallback-accretion model of \citet{Kumar+2008a}, the accretion rate and corresponding jet luminosity can decline approximately as $t^{-3}$ or faster following the prompt emission, as the outer layers of the stellar core collapse and the available mass supply decreases. However, the model also allows shallower luminosity evolution at later times, depending on the continued supply of fallback material and the viscous evolution of the accretion disk. These predictions motivate observational studies of flare luminosity evolution as a means of constraining the long-term activity of the central engine.

The luminosity evolution of X-ray flares has long been regarded as an important observational test of accretion-powered central-engine models. Observational studies have progressively refined this picture. Using a square-flare approximation applied to the flare sample of \citet{Falcone+07}, \citet{Lazzati+2008} found that the average flare luminosity decreases approximately as
$\langle L\rangle \propto t^{-1.5\pm0.16}$
and interpreted this decline as tracing the long-term evolution of the central engine. Extending the analysis to the larger flare sample of \citet{Chincarini+10}, with explicit flare-profile modeling and afterglow subtraction, \citet{Margutti+2011} derived a steeper relation, $\langle L\rangle \propto t^{-2.7\pm0.1}$,
during the interval
$30\lesssim t\lesssim1000$~s. Assuming
$L_{\rm jet}\propto\dot M_{\rm acc}$,
they interpreted this result as evidence for a progressively declining mass accretion rate. At later epochs, \citet{Bernardini+11} found that the peak luminosity evolution becomes considerably flatter, although they argued that this behavior is largely caused by observational incompleteness because only the brightest flares remain detectable above the fading afterglow continuum.

Our results extend these previous observational studies in three important ways. First, our observable differs from those considered previously. Whereas \citet{Margutti+2011} investigated the temporal evolution of the ensemble-averaged flare luminosity and \citet{Bernardini+11} studied the evolution of the peak luminosities of individual flares, we correlate the average luminosity of each modeled flare with its peak time. Although these observables are not identical, they all probe the temporal evolution of the characteristic flare luminosity and therefore provide complementary observational constraints on the long-term activity of the GRB central engine.

Second, our flare luminosities are obtained from a simultaneous decomposition of the complete X-ray light curve, in which the flare emission and the underlying afterglow are modeled self-consistently. As discussed in Appendix~\ref{flare_effects}, this approach indicates that flare emission contributes over a larger fraction of the light curve than previously reported \citep{Chincarini+10, Bernardini+11}, allowing a more consistent separation of the flare and afterglow components and a more self-consistent estimate of the flare luminosities. The broader flare durations inferred from this decomposition naturally increase the fraction of events with $w/t_{\rm pk}>1$ and may contribute to the larger intrinsic dispersion observed for the late--wide flare population.

Third, our analysis provides a different perspective by separating the flare sample into two statistically distinct populations, allowing the luminosity evolution of the early--narrow and late--wide flares to be examined independently.

The early--narrow flares exhibit a significant anti-correlation between average flare luminosity and peak time,

\begin{equation}
L_{\rm avg,f,rest}\propto
t_{\rm pk,rest}^{-1.74^{+0.24}_{-0.24}},
\end{equation}

whereas the late--wide flares are characterized by a much flatter best-fit relation,

\begin{equation}
L_{\rm avg,f,rest}\propto
t_{\rm pk,rest}^{-0.19^{+1.01}_{-0.93}}.
\end{equation}

As a robustness check, we also computed the geometric mean luminosity in logarithmic bins of $t_{\rm pk,rest}$ for the combined flare sample following the methodology of \citet{Margutti+2011} as presented in Figure \ref{fig:scatter_Lavg_tpk} (right). For the combined flare sample, the resulting slope,
$\beta=-1.65^{+0.24}_{-0.20}$,
is fully consistent with that obtained from fitting the individual flare measurements,
$\beta=-1.60\pm0.12$,
demonstrating that the inferred luminosity evolution is robust against whether it is measured from individual flare luminosities or from the geometric mean luminosity in logarithmic time bins. Applying the same procedure to the early--narrow flare population alone yields a consistent slope,
$\beta=-1.70^{+0.69}_{-0.55}$,
showing that the observed correlation persists when the early–narrow population is analyzed separately and is not simply a consequence of combining the two flare populations.

Interestingly, the luminosity evolution of the early–narrow flare population is accompanied by increases in both the flare rise and decay times with peak time, as discussed in Section~\ref{dis:temporal_properties}, while their ratio remains approximately constant. Taken together, these results indicate that progressively later early--narrow flares become less luminous but evolve over longer characteristic timescales, while preserving an approximately constant temporal asymmetry. This suggests an approximate preservation of flare shape during the temporal evolution of the early--narrow population.

The luminosity evolution of the early--narrow population
($\beta=-1.74^{+0.24}_{-0.24}$) is numerically closer to the average luminosity decline reported by \citet{Lazzati+2008}
($\langle L\rangle\propto t^{-1.5\pm0.16}$) than to the steeper
relation found by \citet{Margutti+2011} ($\langle L\rangle\propto t^{-2.7\pm0.1}$). This comparison should,
however, be interpreted with caution because the studies adopt
different luminosity definitions and flare-decomposition methodologies, including differences in the inferred flare durations (Appendix~\ref{flare_effects}). The shallower luminosity decline observed for the early–narrow population also differs from the approximately $t^{-3}$ evolution predicted by \citet{Kumar+2008a} for the rapid post-prompt decline of accretion power, although these quantities do not necessarily trace the same evolutionary regime.

The poorly constrained luminosity evolution of the late–wide population is consistent with a substantially flatter relation, qualitatively resembling the flattening reported by \citet{Bernardini+11} for late-time flares. Although observational selection effects may contribute to this behavior, our identification of a statistically distinct late–wide population raises the possibility that the apparent flattening is not simply the late-time extension of a single flare sequence.

An interesting connection with a specific accretion model is provided by the flare asymmetry. The strongly asymmetric component of the early–narrow population, $t_{\rm rise}/t_{\rm decay}\simeq0.14$, corresponds to $t_{\rm decay}/t_{\rm rise}\simeq7$, close to the approximately model-independent value $t_{\rm decay}/t_{\rm rise}\simeq5.7$ predicted within the disk-fragmentation model of \citet{Dall_Osso+2017}. A similar characteristic asymmetry is also observed among the late–wide flares. Although this similarity is interesting, it does not uniquely identify disk fragmentation as the flare-production mechanism.

If these observational differences are intrinsic rather than dominated by observational selection, one possible interpretation is that the two flare populations do not correspond to different physical mechanisms, but instead represent different evolutionary regimes of the same accretion process. During an earlier, high-accretion-rate phase, accretion may produce relatively luminous and temporally narrow flare episodes. As the accretion flow evolves, gravitational fragmentation, inhomogeneous fallback, viscous evolution, or magnetic regulation may modify the timing and duration of individual accretion episodes \citep{Perna+2006,Proga&Zhang2006,Kumar+2008a}, potentially producing broader and more prolonged or intermittent late-time accretion episodes.

The temporal properties of the two populations provide additional constraints on this interpretation. In accretion-powered models, the observed flare peak time may reflect a combination of the fallback, circularization, and viscous timescales, while the flare duration may also depend on the temporal distribution of the mass supply and the evolution of the accretion flow. The systematically larger width-to-peak-time and rise-to-peak-time ratios of the late–wide flares indicate that their emission and rise phases occupy greater fractions of their characteristic occurrence times than those of the early–narrow population. These differences suggest that the two populations may probe different temporal structures of the accretion flow, potentially associated with variations in the mass-supply history, disk evolution, or the response of the jet to changes in accretion power.

The waiting-time distributions provide an additional constraint on this interpretation. The late–wide second flares are preceded by substantially longer rest-frame intervals between successive flare peaks than the early–narrow second flares, with characteristic waiting times of approximately $2950$~s and $76$~s, respectively, a difference of approximately one and a half orders of magnitude. If these intervals are related to the recurrence timescale of central-engine activity, they may reflect longer intervals between successive episodes of enhanced accretion at later times. Together with the broader temporal structure and flatter best-fit luminosity evolution of the late–wide flares, these longer intervals are consistent with a later and potentially more intermittent phase of central-engine activity. However, the observed waiting times do not necessarily correspond directly to a single physical accretion timescale, and their separation between the two populations is not entirely independent of the peak times used in the population classification. Whether these differences arise from delayed fallback, disk fragmentation, viscous evolution, or another accretion process cannot be distinguished from the present data.

The two populations of X-ray flares, with characteristic rest-frame peak times of approximately 93 and 3090 s, may reflect different accretion episodes associated with distinct regions of a collapsing Wolf–Rayet progenitor. In this scenario, the earlier flares could originate from material in the inner carbon–oxygen-rich layers, whereas the later flares may be associated with fallback from more extended stellar layers, depending on the progenitor structure and angular-momentum distribution \citep{Woosley&Heger2006, Kumar+2008a, Kumar+2008b, Lindner+2010}. For a representative progenitor, material falling from radii of approximately $r\sim10^{10}$ and $10^{11}$~cm has characteristic free-fall times of order 40 and 1000~s, respectively, for enclosed masses of 5 and 10 solar masses. The circularization radius depends on the specific angular momentum of the infalling material and may be substantially smaller than its original radius within the progenitor. Following fallback, the material circularizes and accretes on a viscous timescale, $t_{\rm visc}\sim\alpha^{-1}(H/R)^{-2}\Omega_K^{-1}$. For a newly formed $5 M_\odot$ black hole, with $\alpha\sim0.1$ and $H/R\sim0.3$, viscous timescales of approximately 40 and 2000~s correspond to circularization radii of approximately $4.4\times10^8$ and $6.0\times10^{9}$~cm, respectively. These illustrative combinations of free-fall and viscous timescales are broadly consistent with the two characteristic flare peak times. Moreover, the normalized flare widths, FWHM/$t_{\rm pk}$, are of order unity in both populations, although systematically larger for late–wide flares. This may indicate that viscous evolution contributes to the flare durations, with the relative contributions of fallback and viscous accretion potentially differing between the two populations. While this interpretation remains speculative, it suggests that the two flare populations could trace distinct structural or angular-momentum regions within the progenitor, provided that the intervening layers produce comparatively little episodic accretion.

Taken together, our analysis introduces several new observational constraints on accretion-powered central-engine models. The early–narrow and late–wide flare populations exhibit different normalized temporal properties and waiting-time distributions, together with potentially different luminosity evolution, while the widely reported width–peak-time correlation is primarily associated with the early–narrow population. Moreover, the flare properties are largely independent of the presence of an X-ray plateau \citep{DB+26}, suggesting that the mechanism governing the flare properties is largely independent of that responsible for the plateau phase \citep{DPR22}. Any successful accretion-powered central-engine model should therefore reproduce not only the overall decline of flare luminosity with time, but also the joint distributions of flare peak times, widths, rise and decay times, and waiting times, including the emergence of the two statistically distinct flare populations. The potentially different luminosity evolution of these populations further suggests that a single power-law decline may not adequately characterize the entire flare population. Quantitative comparisons with fallback and disk-evolution models will be needed to determine whether these observational differences can arise from the evolution of a common accretion process or require additional physical ingredients.

The results of this study are also relevant to observations from the recently launched Einstein Probe (EP) and SVOM missions, which provide complementary capabilities for investigating GRB emission. The wide-field soft X-ray observations of EP can probe early X-ray activity that may be missed by Swift-XRT, while SVOM's multiwavelength observations, particularly its optical follow-up capabilities, can provide complementary constraints on the connection between prompt emission and subsequent afterglow activity, as well as the possible relationship between X-ray flares and optical variability. In particular, extending the observational coverage of early X-ray activity and obtaining well-sampled X-ray afterglow light curves may help determine whether the two flare populations identified in this work persist in independent GRB samples.

%%%%%%%%%%%%%%%%%%%%%%%%%%%%%%%%%%%%%%%%%%%%%%%%%%%%%%%%%%
%%%%%%%% Summary and Conclusions
%%%%%%%%%%%%%%%%%%%%%%%%%%%%%%%%%%%%%%%%%%%%%%%%%%%%%%%%%%

\section{Summary and Conclusions}
\label{sec:summary}

We have investigated the temporal and energetic properties of 97 X-ray flares identified in 61 of 89 Swift GRBs with measured redshifts, using the simultaneous flare–afterglow decomposition presented in \citet{DB+25}. Our main observational results can be summarized as follows.

\begin{enumerate}

\item A multivariate Gaussian Mixture Model applied to the joint distribution of the rest-frame flare peak time and width identifies two statistically distinct flare populations. Approximately $80\%$ of the sample belongs to an early--narrow population, while the remaining $20\%$ forms a late--wide population. Although the late--wide population is dominated by second flare episodes (13 out of 19; $68\%$), the majority of all second flares (23 out of 36; $64\%$) belong to the early--narrow population. Thus, flare order alone cannot explain the observed clustering, indicating that the two populations cannot be explained solely by the temporal sequence of flare episodes. The classification is also robust against cosmological time dilation, with approximately 93\% of the flares retaining their cluster assignments when the GMM analysis is performed in the observer frame rather than the rest frame.

\item The two populations exhibit significantly different temporal properties. Early--narrow flares have systematically smaller normalized rise times and widths with flare peak time, whereas late--wide flares evolve over substantially longer characteristic timescales. The rise times increase significantly with flare peak time in both populations, while the decay times show a significant increase for the early–narrow flares and a weaker trend for the late–wide flares. The absence of significant evolution in the temporal asymmetry suggests that the flare profiles remain approximately self-similar over a broad range of peak times.

\item The early--narrow flare population exhibits a possible bimodal distribution of the temporal asymmetry ratio $t_{\rm rise}/t_{\rm decay}$. The first component of the bimodality is characterized by a stronger temporal asymmetry, with decay times approximately seven times longer than rise times. Interestingly, this asymmetry is consistent with the dominant asymmetry observed in the late--wide flare population. This bimodality persists after excluding potentially prompt contaminated flares ($T_{90}>t_{\rm pk}$), indicating that the observed structure is not driven by prompt-emission contamination.

\item The widely reported width--peak-time correlation does not describe the two flare populations equally well. The early--narrow flares retain a strong correlation with a regression slope consistent with unity, whereas the late--wide flares show only a moderate, statistically insignificant correlation and a substantially less constrained scaling. The strong global correlation therefore reflects both the intrinsic scaling of the early--narrow population and the separation of the two populations in the $w$--$t_{\rm pk}$ plane.

\item Our simultaneous modeling of the flare and afterglow components indicates that flare emission contributes to the observed X-ray light curve over a longer interval than previously recognized, naturally recovering broader flare profiles and increasing the fraction of events with $w/t_{\rm pk}>1$. This result provides a more self-consistent separation of the flare and afterglow components and may partly explain the larger intrinsic dispersion observed among the late--wide flares.

\item The flare energetics are only weakly linked to the temporal properties of the two populations. The isotropic flare-energy distributions are statistically consistent between the early--narrow and late--wide populations ($p=0.22$), despite the lower average energy of the late--wide flares. No statistically significant correlation is found between flare width and isotropic energy in either population, while only a weak anti-correlation between flare peak time and isotropic energy is found for the early--narrow flares. 

\item In contrast, the luminosity–peak-time relation is more strongly constrained for the early–narrow population than for the late–wide population. The early–narrow flares exhibit a significant anti-correlation between the average flare luminosity and flare peak time, with 
$
L_{\rm avg,f,rest}\propto t_{\rm pk,rest}^{-1.74\pm0.24}
$
, whereas no statistically significant correlation is detected for the late–wide population ($\beta=-0.19^{+1.01}_{-0.93}$). For the combined sample, the slope obtained from fitting individual flare measurements ($\beta=-1.60\pm0.12$) is consistent with that obtained independently from the binned geometric mean luminosities ($\beta=-1.65^{+0.24}_{-0.20}$). These results suggest that the overall luminosity decline reported in previous studies is primarily associated with the early–narrow flare population, while the late–wide flares contribute additional scatter.

\item The two flare populations exhibit different waiting-time distributions among second flares, with substantially longer waiting times for late–wide flares ($\sim 2950~s$; 13 events) than for early–narrow flares ($\sim 76~s$; 21 events). Together with their broader temporal profiles and weaker observed luminosity–peak-time relation, this suggests that late–wide flares may trace a later, potentially more intermittent phase of central-engine activity.

\end{enumerate}

Taken together with our previous finding that the temporal properties of X-ray flares are independent of the presence of an X-ray plateau \citep{DB+26}, the results presented here further support the interpretation that X-ray flares originate from prolonged central-engine activity rather than from variability in the external shock.

The distinct temporal properties of the early–narrow and late–wide flare populations, particularly their normalized flare widths ($w/t_{\rm pk}$), temporal asymmetries, and waiting-time distributions, together with their observed luminosity evolution, provide new observational constraints on accretion-powered central-engine models. In particular, the significant luminosity decline of the early–narrow flares is qualitatively consistent with a decreasing accretion-powered energy supply, whereas the comparatively shallow but poorly constrained luminosity evolution of the late–wide flares may indicate a different regime of prolonged accretion activity. Although observational selection effects cannot yet be excluded, these findings motivate models that can reproduce both flare populations within the temporal evolution of the central engine.

%%%%%%%%%%%%%%%%%%%%%%%%%%%%%%%%%%%%%%%%%%%%%%%%%%%%%%%%%%
%%%%%%%% Acknowledgments
%%%%%%%%%%%%%%%%%%%%%%%%%%%%%%%%%%%%%%%%%%%%%%%%%%%%%%%%%%
\begin{acknowledgments}
This work made use of data supplied by the UK Swift Science Data Center at the University of Leicester.
AP acknowledges support from the European Union via ERC consolidating grant $\sharp$773062 (acronym O.M.J.) and from the Israel Space Agency via grant number 6766.

\end{acknowledgments}

\vspace{5mm}
\facilities{\textit{Swift}-XRT}
%\software{MultiNest}

%%%%%%%%%%%%%%%%%%%%%%%%%%%%%%%%%%%%%%%%%%%%%%%%%%%%%%%%%%
%%%%%%%% Bibliography
%%%%%%%%%%%%%%%%%%%%%%%%%%%%%%%%%%%%%%%%%%%%%%%%%%%%%%%%%%
\bibliography{biblio_new}{}
\bibliographystyle{aasjournal}

%%%%%%%%%%%%%%%%%%%%%%%%%%%%%%%%%%%%%%%%%%%%%%%%%%%%%%%%%%
%%%%%%%% Appendix
%%%%%%%%%%%%%%%%%%%%%%%%%%%%%%%%%%%%%%%%%%%%%%%%%%%%%%%%%%
\appendix
\restartappendixnumbering  % Resets numbering and adds the "A" prefix

\section{Robustness of the Flare Classification and Temporal Properties}
\label{sec:robustness_classification}
The identification of two statistically distinct flare populations relies on both the multivariate GMM classification and the temporal properties derived from the simultaneous flare--afterglow decomposition. It is therefore important to verify that these results are not driven by observational or methodological effects. In this Appendix section, we examine the influence of cosmological time dilation, the treatment of the underlying afterglow, and possible prompt-emission contamination. We find that none of these effects alters the principal conclusions of this work, supporting the interpretation that the early--narrow and late--wide flare populations represent intrinsic features of the observed GRB flare sample.

\subsection{GMM application to the flare peak time and width in the observer frame}
\label{GMM_tpk_w_observer_frame}
To investigate whether the observed bimodality could be produced by cosmological time dilation, we first applied the multivariate GMM to the flare peak time and width in the observer frame. The analysis identifies two well-separated populations: an early--narrow component (centered at $\log_{10}(t_{\rm pk})\simeq2.34$ and $\log_{10}(w)\simeq2.24$) and a late--wide component (centered at $\log_{10}(t_{\rm pk})\simeq3.62$ and $\log_{10}(w)\simeq3.76$). We then repeated the same analysis using the rest-frame quantities as presented in Section \ref{sec:narrow_wide_flares}.
The rest-frame analysis again identifies two statistically distinct populations with properties very similar to those obtained in the observer frame, demonstrating that the early--narrow and late--wide classification is largely intrinsic rather than an artifact of cosmological time dilation.

\begin{figure}
    \centering
    \includegraphics[width=\linewidth]{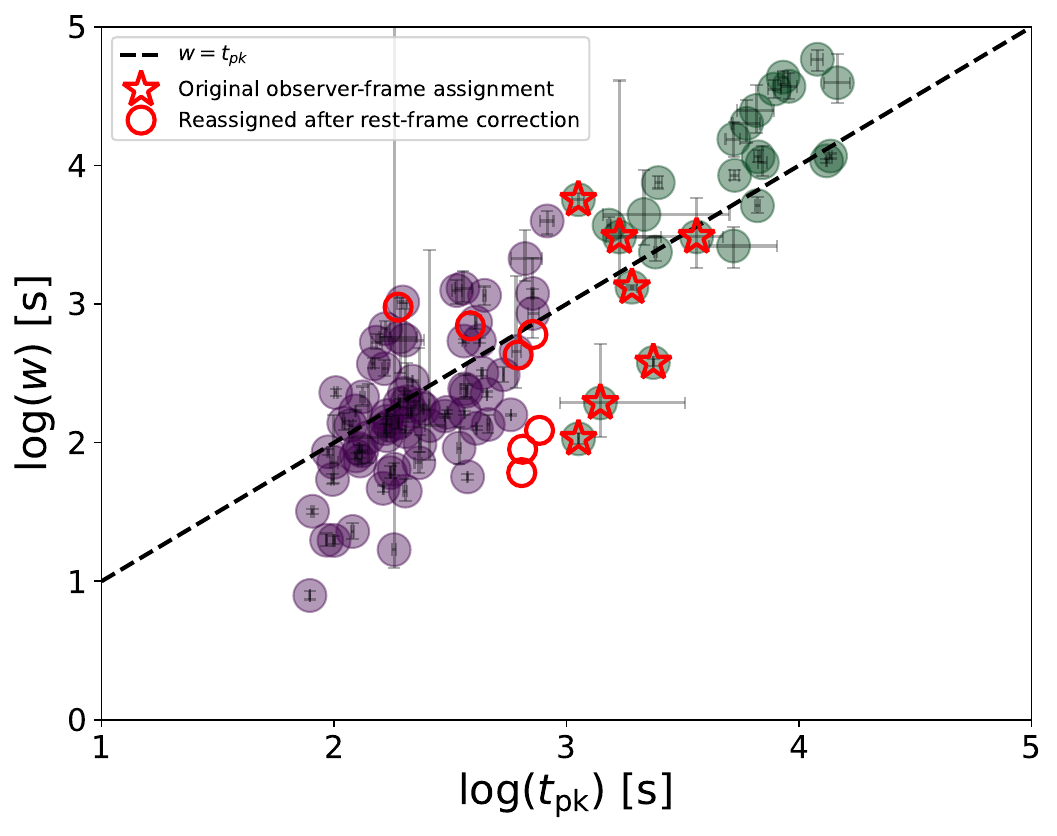}
    \caption{Comparison of the multivariate GMM classification in the observer and rest frames. The flare peak time ($t_{\rm pk}$) and flare width ($w$) are shown in logarithmic space. The purple and green points represent flares assigned to the early--narrow (Cluster~0) and late--wide (Cluster~1) populations, respectively, according to the observer-frame GMM classification. The error bars correspond to $1\sigma$ uncertainties in the linear scale. The dashed line indicates $w=t_{\rm pk}$. The red star symbols mark the original observer-frame positions of the seven flares whose cluster assignment changes after correcting to the rest frame, while the open red circles indicate their corresponding positions in the rest frame. The small displacement of these seven flares demonstrates that cosmological time dilation has only a minor effect on the clustering, with the vast majority of flares retaining the same classification.}
    \label{fig:restframe_cluster_changes}
\end{figure}

Only seven out of the 97 flares ($\sim7\%$ of the sample) change their cluster assignment after correcting to the rest frame (see Figure~\ref{fig:restframe_cluster_changes}). All of these flares lie close to the boundary between the two GMM components and are reassigned from the late--wide to the early--narrow population. The remaining $\sim93\%$ of the sample retains the same classification, indicating that cosmological time dilation has only a minor effect on the clustering and cannot account for the observed bimodality.
These results justify the adoption of the rest-frame GMM classification throughout the remainder of this work.

%%%%%%%%%%%%%%%%%%%%%%%%%%%%%%%%%%%%%%%%%%%%%%%%%%%%%%%%%
%%%%%%% Figure: $w$ vs. $t_{\rm pk}$ correlation in the observer and rest frames
%%%%%%%%%%%%%%%%%%%%%%%%%%%%%%%%%%%%%%%%%%%%%%%%%%%%%%%%%

%%%%%%%%%%%%%%%%%%%%%%%%%%%%%%%%%%%%%%%%%%%%%%%%%%%%%%%%%
%%%%%%%
%%%%%%%%%%%%%%%%%%%%%%%%%%%%%%%%%%%%%%%%%%%%%%%%%%%%%%%%%

\subsection{Wider flare width: The influence of flares on the underlying afterglow}
\label{flare_effects}

One of the advantages of our fitting approach is that it explicitly models the underlying afterglow emission simultaneously with the flare components using physically motivated synchrotron models. This allows us to place stronger constraints on the underlying afterglow and to identify flare emission that remains significant over a longer time interval than is typically recovered using simple broken power-law descriptions of the afterglow. In particular, we find that %almost 
10 out of the 78 early--narrow and 10 out of the 19 late--wide flares extend well beyond the temporal range over which flare emission has generally been considered in previous studies. Such broad flares can be difficult to distinguish from the underlying continuum when the afterglow is approximated by empirical power-law segments, since part of the flare emission may instead be absorbed into the continuum model.

This issue is also reflected in our previous analysis \citep{DB+25}, where 11 GRBs were excluded from the original sample of 100 bursts because the fitted flare widths became extremely large, in some cases extending beyond $10^{5}$~s and covering more than half of the observed temporal range. At the time, these events were conservatively excluded because the flare component dominated nearly the entire light curve, preventing a reliable decomposition of the flare and afterglow contributions. However, the present analysis suggests an alternative interpretation. Rather than representing purely pathological fits, at least some of these events may indicate that the influence of flare emission on the observed afterglow persists much longer than previously assumed. If this interpretation is correct, then late-time flare emission may remain detectable as a low-level excess above the underlying synchrotron afterglow, emphasizing the importance of simultaneously modeling both components when deriving the physical properties of GRB afterglows.

\subsection{Prompt emission contamination on the $t_{\rm rise}/t_{\rm decay}$ distributions}
\label{promt_effect}

As shown in Figure~\ref{fig:flare-rise_decay_ratio} (top-left), we further investigated the distribution of the flare asymmetry ratio, $t_{\rm rise}/t_{\rm decay}$, for the two GMM flare populations. This ratio provides a measure of the temporal asymmetry of the flare profile, with smaller values corresponding to faster rise phases and longer decay phases.

To assess whether the observed bimodal structure could be affected by contamination from the prompt emission, we identified flares satisfying the condition $T_{90}>t_{\rm pk}$, for which high-latitude emission from the prompt phase may still contribute to the observed X-ray flux. We identified seven GRBs (210905A, 190719C, 180624A, 170519A, 160425A, 160131A, and 150818A) containing a total of 12 flares potentially affected by prompt-emission contamination. However, the second flares of GRB~170519A and GRB~160131A satisfy $t_{\rm pk}>T_{90}$ and are therefore unlikely to be contaminated. Consequently, only 10 flares were excluded from the analysis. All of the excluded flares belong to Cluster~0 (early--narrow flares), while none belong to Cluster~1 (late--wide flares).

After removing these potentially contaminated flares (see Figure~\ref{fig:flare-rise_decay_ratio}, bottom-left), we repeated the univariate two-component GMM analysis separately for the early--narrow and late--wide flare populations in order to test the robustness of the inferred bimodal structure.

For the early--narrow flares (Cluster~0), the GMM identifies two components:
\begin{itemize}
\item \textbf{Peak 1:} $t_{\rm rise}/t_{\rm decay}=0.14~ (\sigma = 0.03)$, representing 25\% of the sample.
\item \textbf{Peak 2:} $t_{\rm rise}/t_{\rm decay}=0.38 ~(\sigma = 0.12)$, representing 75\% of the sample.
\end{itemize}
The first peak corresponds to highly asymmetric FRED-like flares, with decay times approximately seven times longer than rise times, while the second peak corresponds to less asymmetric flares, with decay times approximately three times longer than rise times.

For completeness, we also applied the same two-component GMM to the late--wide flares (Cluster~1):
\begin{itemize}
\item \textbf{Peak 1:} $t_{\rm rise}/t_{\rm decay}=0.14~(\sigma =0.02)$, representing 58\% of the sample.
\item \textbf{Peak 2:} $t_{\rm rise}/t_{\rm decay}=0.32~(\sigma =0.06)$, representing 42\% of the sample.
\end{itemize}

However, because the late--wide sample contains only 19 flares and its distribution is visually consistent with a single broad peak accompanied by a tail toward higher values of $t_{\rm rise}/t_{\rm decay}$, this apparent two-component structure should be interpreted with caution. Overall, both flare populations are characterized by $t_{\rm rise}/t_{\rm decay}<1$, confirming that X-ray flares
are generally asymmetric, with rise times shorter than their decay times.

Excluding the potentially prompt-contaminated flares does not significantly modify the $t_{\rm rise}/t_{\rm decay}$ distribution of Cluster~0. In particular, the bimodal structure remains present, with nearly identical GMM component centers before and after the exclusion as discussed in Section \ref{dis:temporal_properties}. This demonstrates that the observed bimodality is robust against prompt-emission contamination and is therefore likely an intrinsic characteristic of the early--narrow flare population.

\end{document}